# A parallel multistate framework for atomistic non-equilibrium reaction dynamics of solutes in strongly interacting organic solvents

David R. Glowacki,[1,2,3,4]* Andrew J. Orr-Ewing,[1] and Jeremy N. Harvey[5]

[1]*School of Chemistry, University of Bristol, Bristol, BS8 1TS, UK*
[2]*Department of Computer Science, University of Bristol, BS8 1UB, UK*
[3]*PULSE Institute and Department of Chemistry, Stanford University, Stanford, CA 94305, USA*
[4]*SLAC National Accelerator Laboratory, Menlo Park, California 94025, USA*
[5]*Department of Chemistry, KU Leuven, Celestijnenlaan 200F, B-3001 Heverlee*

*drglowacki@gmail.com

**Abstract**

We describe a parallelized linear-scaling computational framework developed to implement arbitrarily large multi-state empirical valence bond (MS-EVB) calculations within CHARMM. Forces are obtained using the Hellman-Feynmann relationship, giving continuous gradients, and excellent energy conservation. Utilizing multi-dimensional Gaussian coupling elements fit to CCSD(T)-F12 electronic structure theory, we built a 64-state MS-EVB model designed to study the $F + CD_3CN \rightarrow DF + CD_2CN$ reaction in $CD_3CN$ solvent. This approach allows us to build a reactive potential energy surface (PES) whose balanced accuracy and efficiency considerably surpass what we could achieve otherwise. We use our PES to run MD simulations, and examine a range of transient observables which follow in the wake of the reactive event, including transient spectroscopy of the DF vibrational band, time dependent profiles of vibrationally excited DF in $CD_3CN$ solvent, and relaxation rates for energy flow from DF into the solvent, all of which agree well with experimental observations. Immediately following deuterium abstraction, the nascent DF finds itself in a non-equilibrium regime in two different respects: (1) it is highly vibrationally excited, with ~23 kcal mol$^{-1}$ localized in the stretch; and (2) it is not yet Hydrogen-bonded to $CD_3CN$ solvent molecules, with a microsolvation environment intermediate between the non-interacting gas-phase limit and the solution-phase equilibrium limit. Vibrational relaxation of the nascent DF results in a spectral blue shift, while relaxation of the microsolvation environment results in a red shift. These two competing effects result in a post-reaction relaxation profile which distinct from what is observed when vibrational excitation of DF occurs within an equilibrium microsolvation environment. The theoretical and parallel software framework presented in this paper should be more broadly applicable to a range of complex reactive systems.

# Introduction

The accurate simulation of reaction dynamics in condensed phase environments remains a formidable challenge within the field of computational and theoretical chemistry.[1-5] There are a wide variety of reasons for this. First and foremost, the accurate treatment of electron correlation in large systems that include solvent/solute coupling remains a significant challenge, even within the Born-Oppenheimer approximation. Consequently, it remains a substantial challenge to develop a potential energy surface which is accurate enough to interpret experiments and also adequately efficient to provide statistically converged results on a reasonable time scale. Even with an accurate and efficient potential, there are further complications which arise according to the fact that experimental dynamic observables ultimately derive from the quantum mechanical properties of molecules, but efficient methods for carrying out full quantum mechanical dynamics simulations of condensed phase systems are not generally possible. Therefore, most approaches utilize classical mechanics, often invoking correction factors to bring the classical simulations into agreement with the known quantum mechanical results for simple model systems.[6,7] In the vicinity of equilibrium, there are a number of approaches available which have been well tested;[6,8] however, far from equilibrium (e.g., at the sorts of energies required to facilitate chemical reactions, where many quanta of excitation are localized in a particular vibrational mode), then the approaches for mapping classical results onto quantum mechanical observables are far less developed.[9]

Conventional approaches utilizing parameterized force-fields continue to play an important role in computational investigations of the condensed phase.[10] Such force-fields have enjoyed considerable success answering a range of questions spanning biochemical systems, materials science, and solvent dynamics, but they also have a number of well-known shortcomings. From the perspective of the work presented in this paper, the most significant shortcoming arises from the fact that force-field parameterization schemes tend to focus on equilibrium properties in the vicinity of energy minima where anharmonicity is very small. As a result, they tend to do an excellent job in the vicinity of stationary point minima. However, if the system strays from the minima, then they often give results which are unable to accurately answer a range of experimentally relevant questions. For example, bond breaking and forming usually occur far from the minima, requiring several quanta of vibrational excitation in a particular bond.[11] Similarly, relaxation dynamics often occur in regions of phase space that are far from the parameterized minima. Detailed studies of phenomena that occur in regions of configuration space which are far from the minima (e.g., reaction and relaxation dynamics) therefore have an important role to play in refining force-field type approaches.

Advances in experimental techniques provide increasingly detailed measurements in condensed phase systems, providing excellent tests of computational approaches for investigating condensed phase reaction dynamics.[12-17] The last decade or so has seen a number of attempts by a range of workers to efficiently and accurately simulate chemical reaction dynamics in condensed-phase environments using multi-state molecular mechanics type approaches.[18-30] These sorts of approaches represent the important molecular configurations using diabatic valence bond states, and a wide range of schemes for calculating the coupling between the diabatic states.[22, 31-34] So far, such approaches have largely been confined to simulation of enzymatic chemical reactions, and proton transfer in aqueous environments. Despite well-characterized problems with energy conservation that arise from using a truncated or moving basis set of diabatic states, such approaches have seen heavy use, no doubt linked to their efficiency.

In recent work we have investigated reaction dynamics of solutes in weakly coupled organic solvents:[3, 35] namely, $CN + C_6H_{12} \rightarrow HCN + C_6H_{11}$ in dichloromethane solvent,[36-38] and $CN + C_4H_8O \rightarrow HCN + C_4H_7O$ in tetrahydrofuran ($C_4H_8O$) solvent.[39] In both cases, we carried out reaction dynamics studies focussed on unravelling the mechanisms by which solvent interactions relaxed the vibrationally hot products formed during the abstraction process. These studies exploited some simple EVB models along with a recently developed rare event acceleration algorithm which is able to generate statistically meaningful non-equilibrium dynamics in the post transition state region. To the best of our knowledge, these theoretical investigations of non-equilibrium solution phase bimolecular reaction dynamics were amongst the first of their kind. Through detailed comparison with ultrafast spectroscopy experiments, we were able to reveal microscopic detail into how the relaxation dynamics which follow a chemical reaction differ from the relaxation dynamics which follow vibrational relaxation in an equilibrated microsolvation environment.[1]

Building on our earlier studies,[36-38] the work described in this paper carries out simulations of F atom abstraction reactions in $CD_3CN$ solvent to give $DF + CD_2CN$, in which the solute/solvent interactions are considerably stronger with considerably more anharmonic coupling in dynamically relevant regions of the phase space. An accurate treatment of the dynamics thus requires a simulation framework more complicated than that utilized previously. Specifically, we used a locally modified version of the CHARMM program suite[40] to build a parallel 64 state EVB model from Gaussian coupling elements whose parameters were fit against CCSD(T)-F12 electronic structure theory calculations (we note that we have recently implemented a nearly identical MS-EVB parallelization of TINKER[27]). The net result is a general parallel simulation framework which is able to describe the reaction dynamics of the abstraction event, as well as the subsequent relaxation dynamics, because it includes an accurate description of the abstraction potential as well as interactions between the solvent and the nascent DF. The latter was achieved by including ionic diabatic states which permit the nascent DF to undergo transient deuteron transfer to all the solvent molecules within the simulation, similar to approaches recently utilized to investigate the spectral shifts of proton shifts associated with hydrogen bonded complexes.[29]

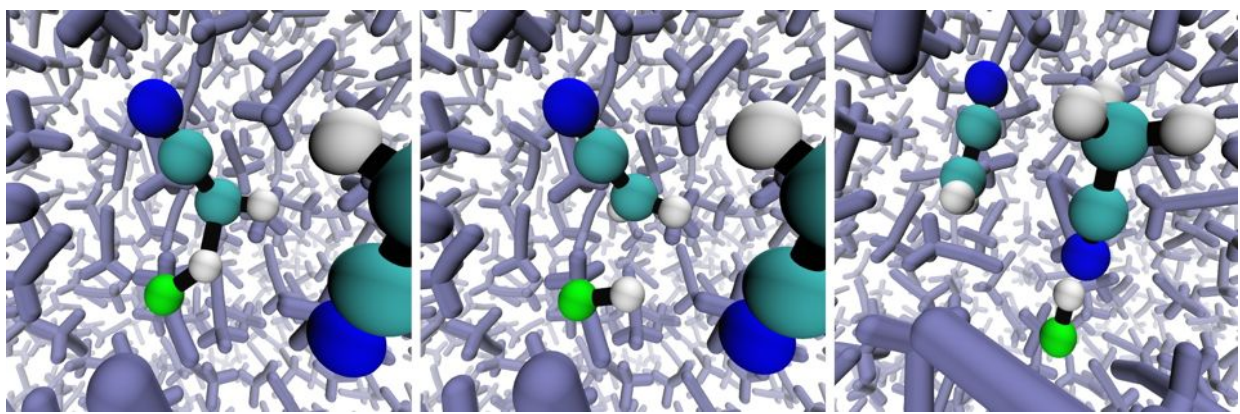

Figure 1: Snapshots from molecular dynamics simulations. The left hand panel shows the Flourine radical embedded in $CD_3CN$ solvent prior to abstraction, the middle panel shows the moment immediately following abstraction to form DF, and the right hand panel shows the diatomic DF product hydrogen bonded to one of the solvent molecules

Figure 1 shows a schematic of the reaction to which we apply our computational framework, namely

$$F + CD_3CN \text{ (in } CD_3CN) \rightarrow DF + CD_2CN \text{ (in } CD_3CN) \qquad (R1)$$

which may be usefully decomposed into the following elementary steps:

$$F + CD_3CN \text{ (in } CD_3CN) \rightarrow DF^* + CD_2CN \text{ (in } CD_3CN) \qquad (R1a)$$
$$DF^* \text{ (in } CD_3CN) \rightarrow DF \text{ (in } CD_3CN) \qquad (R1b)$$

The simulations described in this paper reveal that the transient spectral profiles of the nascent DF* produced in (R1a) and relaxed in (R1b) result from a set of different physical effects which are closely coupled and occur on very similar timescales, involving: (1) Vibrational anharmonicity which depends on both the force field and the energy content of the DF solute; (2) Time-dependent solute/solvent spectral overlap; (3) The magnitude of the solute/solvent coupling; (4) The relaxation rate of DF within an equilibrium $CD_3CN$ micro-solvation environment; and (5) the timescale at which the time-dependent microsolvation environment of the nascent DF approaches equilibrium.

In organizing this paper, we have decided to focus on each of the elements described above, in order to decompose the distinct effects responsible for the time dependent DF spectra obtained our simulations. This paper is broadly divided into two different sections. The first section focuses on the potential energy methods which we utilized to model (R1a) and (R1b). Here we outline electronic structure theory calculations [density function theory (DFT) and coupled cluster with singles, doubles, and perturbative triples (CCSD(T))] carried out to model two important regions of the PES: geometries along the abstraction path, and also complexes between the nascent DF and $CD_3CN$ solvent molecules. These electronic structure theory calculations were used as a test set for fitting solute/solvent coupling in a 64-state MPI-parallelized MS-EVB computational model. This approach allows us to accurately treat both the abstraction dynamics and the coupling between the nascent DF solute and the $CD_3CN$ solvent molecules within the dynamically important energy range.

The second section focuses on the dynamics simulations which we carried out to describe (R1b). First, we describe gas phase simulations of DF as a Morse oscillator, outlining the energy dependent spectral shifts observed for this simple model. Second, we outline the energy relaxation results obtained when an excited DF Morse oscillator is solvated in bath of $CD_3CN$ solvent. Here, we pay particular attention to how the DF relaxation timescales depend on solute/solvent spectral overlap, and the strength of solvent/solute coupling. Finally, we describe reactive simulations of (R1) in its entirety, showing how the observed time-dependent spectra depend on the effects outlined previously. The microscopic picture that emerges from these simulations reveals a complicated time-dependent spectral profile of the nascent DF, which results from two opposing effects: there is a fast red shift that occurs as the nascent DF relaxes to form complexes with neighboring solvent molecules, and there is also a competing blue shift as DF loses its vibrational energy to the solvent.

## Potential Energy Surfaces

### *Electronic Structure calculations*

Non-equilibrium dynamics simulations are notoriously sensitive to the details of the potential energy surface. For example, several workers, going back to Polanyi, have shown that product energy deposition depends very sensitively on the shape of the potential energy surface.[41, 42] Similarly, it has also been extensively shown that vibrational energy relaxation rates of simple solutes depend sensitively on the coupling of anharmonic modes in the solute and solvent potential energy surface.[43-45] Along these lines, the first part of this study involved obtaining an accurate representation of these two important regions of the F + $CD_3CN$ potential energy surface – i.e., at geometries sampled in the vicinity of: (1) the abstraction region of the potential energy surface, which determines product energy deposition; and (2) the subsequent complexes which may be formed upon conclusion of the abstraction reaction ($CD_2CN\cdots DF$ and $CD_3CN\cdots DF$) which impacts the relaxation dynamics of the abstraction products.

To determine an accurate set of geometries and their corresponding energies in these different regions of the PES, we utilized the following procedure: Structures for the separated reactants and products, the $CD_3CN \bullet F$ reactant complex, the $CD_2CN$—DF product complex, the solvent complex $CD_3CN$—DF and the F---H-CD2CN abstraction TS were optimized using density functional theory, with the M06-2X functional and the 6-311+G(d,p) basis set within the Gaussian program suite.[46] Frequencies were computed to generate zero-point energy corrections. Additional structures were generated at points lying close to the minimum energy path for hydrogen abstraction. These structures were obtained by optimizing the structure of acetonitrile while holding the C-H bond length fixed at a set of values ranging from 0.8 Å to 2.8 Å, in steps of 0.1 Å. This was done using the same M06-2X functional and basis set, and with careful checking for lower-energy unrestricted solutions at larger C-H distances. Then the fluorine atom was positioned collinearly along the C-H bond direction at a set of distances. The structure and energy of the point located at C-H and H-F distances close to that found in the optimized TS was similar to that of the TS structure itself. Additional structures corresponding to distorted $CD_3CN$—DF species were obtained in two ways. First, this complex was reoptimized while holding the H—F distance frozen at values between 0.80 and 1.1 Å, in steps of 0.05 Å. Inspection of these structures showed that the internal structure of the $CD_3CN$ moiety barely changed with respect to equilibrium. Hence additional structures were obtained by varying the N-H and H-F distances while holding other coordinates fixed at those for the equilibrium $CD_3CN$—DF complex. Single-point energies at all these structures were computed using the CCSD(T) method with explicit treatment of electron-electron correlation using the F12-b ansatz within the MOLPRO program suite.[47-49] The cc-pVTZ basis set was used for H and C atoms, and the aug-cc-pVTZ basis for N and F. Appropriate auxiliary basis sets from the aug-cc-pVTZ family were used on all atoms. In what follows, we report relative energies based on these calculations, which is what we used to parameterize reactive MD potential energy surfaces.

Table 1 shows zero-point corrected results obtained using the electronic structure methods discussed above. Our main motivation for choosing the M06-2X functional is that it returns a TS energy similar to that obtained with coupled-cluster theory. Table 1 shows that the fluorine atom forms a weak pre-reaction complex with the lone pair of acetonitrile. The reaction has a low barrier, which is decreased by zero-point energy correction. The TS is near linear, with a C—D—F angle of 162°, and is early, with C—D and D—F distances of 1.131 Å and 1.493 Å respectively. Both the product radical and $CD_3CN$ form strong hydrogen bonds with DF. This leads to a slight elongation of the DF bond, from 0.918 Å in isolated DF to 0.932 Å in the complex. The CCSD(T) calculations predict that this complex is bound by 9.7 kcal mol$^{-1}$. As will be shown later, the existence of strong hydrogren bonds between nascent DF and the cyano moiety of the $CD_3CN$ solvent molecules is important to the dynamics results. Table 1 shows results obtained from the electronic structure theory calculations, including stationary point energies and vibrational frequencies.

Table 1: Calculated potential energies (in kcal mol$^{-1}$) for species involved in the F + $CD_3CN$ reaction, and subsequent interaction of DF with solvent molecules. [a] Energies calculated at the M06-2X structures; [b] Energy relative to $CD_3CN$ + DF.

| Species | $E_{rel}$(M06-2X) | $E_{rel}$(M06-2X+zpe) | $E_{rel}$(CCSD(T)-F12)[a] |
|---|---|---|---|
| F + $CD_3CN$ | 0.0 | 0.0 | 0.0 |
| $CD_3CN$•F | −3.7 | −3.3 | −2.1 |
| TS | 1.3 | −0.7 | 2.5 |
| $CD_2CN$—DF | −43.1 | −44.3 | −45.1 |
| $CD_2CN$ + DF | −34.3 | −37.3 | −36.8 |
| $CD_3CN$—DF[b] | −9.6 | −7.7 | −9.1 |

Table 2: Calculated M062x/6-311+G(d,p) frequencies (in cm$^{-1}$) for species involved in the F + $CD_3CN$ reaction, and subsequent interaction of DF with solvent molecules.

| Species | Vibrational frequencies |
|---|---|
| $CD_3CN$ | 352, 354, 844, 864, 864, 1062, 1063, 1134, 2208, 2339, 2341, 2423 |
| DF | 3050 |
| $CD_2CN$ | 344, 399, 540, 853, 925, 1168, 2215, 2312, 2461 |
| $CD_3CN$—DF | 41, 51, 180, 346, 399, 457, 460, 548, 855, 930, 1167, 2250, 2312, 2464, 2831 |
| $CD_2CN$—DF | 46, 48, 184, 357, 358, 479, 479, 853, 866, 866, 1059, 1060, 1135, 2209, 2342, 2344, 2446, 2799 |

## *Multi-State EVB calculations*

Running direct dynamics using the CCSD(T) methods described above would have been prohibitively expensive. It would probably have been possible to exploit recent advances in computational efficiency to perform DFT calculations on this system,[50, 51] but the cost would have nevertheless been significant, and would not have yielded satisfactory statistics for interpreting the dynamics. Consequently, we sought other means to develop an accurate analytical representation of the electronic structure results discussed above. Building on from previous work, we developed an EVB model fit to the CCSD(T) results. In the EVB approach, basis functions that effectively correspond to different molecular valence states are used to formulate a Hamiltonian matrix $\mathbf{H}(\mathbf{q})$ to describe the molecular system energy. The diagonal elements of this matrix, $V_i(\mathbf{q})$, correspond to the molecular mechanics energy of a particular valence state specified by a particular connectivity. The off-diagonal elements $H_{ij}(\mathbf{q})$ then describe how strongly different molecular configurations are coupled to one another. Similar to the approach we have taken in previous work, the off-diagonal elements are a function of the system coordinates $\mathbf{q}$.

To reproduce the electronic structure results described above, and to account for the fact that nascent DF formed following D abstraction from CD$_3$CN is able to form complexes with any of $n$ solvent molecules included within the simulation, we required four different types of valence states. These are shown in Figure 2, for the simplest illustrative case, with $n = 2$ solvent molecules. State 1 shows a Fluorine radical nestled between two distinct solvent molecules. In principle, the Fluorine could abstract any of 3 Deuterium atoms from any of the $n$ solvent molecules, resulting in $3n$ possible abstraction processes. In practice, to reduce the computational expense of the simulations, the Fluorine is allowed to abstract a single D atom from a particular nearby solvent molecule. The other $n - 1$ solvent molecules are not reactive, but they nevertheless interact with the reacting system as it progresses along the reaction coordinate from reactants to products.[1] State 2 corresponds to the products formed following the allowed abstraction process. The nascent DF may subsequently form a post-reaction complex – with either its radical coproduct, or any of the other $(n - 1)$ solvent molecules within the simulation. State 3 corresponds to Deuterium transfer from DF to the Nitrogen atom of its co-radical product, and state 4 corresponds to proton transfer from DF to the Nitrogen atom of the other solvent molecule. In general, there are $(n - 1)$ replicas of state 4 (allowing DF to transfer a proton to every non-reactive solvent molecule in the simulation). In the case of Fig 2, where $n = 2$, there is only one such interaction.

---

[1] The decision to allow only one reactive deuterium is not an inherent limitation of our approach. By adding more states, it would have been possible to make it so that the Fluorine could abstract any given hydrogen atom by adding more states, albeit at a larger computational expense. Because the emphasis in this study is on the post transition-state dynamics that follow D abstraction rather than association kinetics, the error arising from this simplification is relatively minor. As shown in the discussion of the dynamics results, a far more significant source of error arises if one neglects the coupling between the nascent DF and its neighboring solvent molecules.

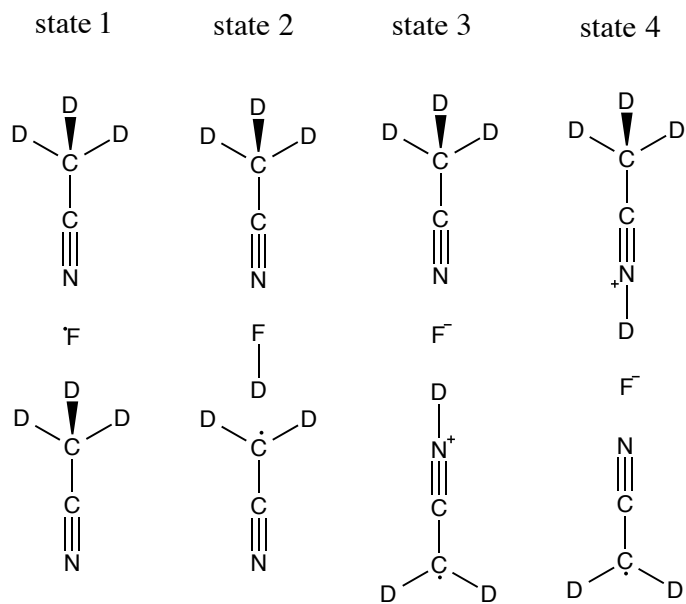

Figure 2: schematic of the diabatic states utilized in our model. For simplicity, we have shown the states that arise for a Fluorine radical embedded in a solvent bath composed of only two CD$_3$CN solvent molecules. For the simulations detailed in the text, F was embedded in $n = 62$ solvent molecules

In our previous work on CN + C$_6$H$_{12}$ abstraction reactions, a sufficiently accurate Hamiltonian matrix was obtained by utilizing two states (i.e., a 2 × 2 matrix). For reasons discussed in detail below, the 2 × 2 approach is inadequate for the system shown in Fig 2.[2] The treatment required for accurate simulation in this work is rather different: the Hamiltonian matrix for a Fluorine radical embedded in $n$ solvent molecules has a dimension of $(n + 2) \times (n + 2)$, with the following structure:

$$\mathbf{H} = \begin{bmatrix} V_1 + \varepsilon_1 & H_{12} & 0 & 0 & \cdots & 0 \\ 0 & V_2 + \varepsilon_2 & H_{23} & H_{24} & \cdots & H_{24} \\ 0 & H_{23} & V_3 + \varepsilon_3 & 0 & \cdots & 0 \\ 0 & H_{24} & 0 & V_4 + \varepsilon_4 & \cdots & 0 \\ \vdots & \vdots & \vdots & \vdots & \ddots & 0 \\ 0 & H_{24} & 0 & 0 & 0 & V_{n+2} + \varepsilon_{n+2} \end{bmatrix} \quad (1)$$

---

[2] For DF solute in CD$_3$CN solvent, the strong coupling between states 2, 3, and 4 in Fig 2 is not well-captured by standard force fields. In our previous work on HCN in CH$_2$Cl$_2$, the solute/solvent coupling was considerably weaker, and reasonably well captured by standard force fields.

where the diagonal elements $V_1$, $V_2$, and $V_3$ respectively correspond to the energies of states 1, 2, and 3 in Fig 2. Diagonal elements with indices spanning $V_4$ to $V_{n+2}$ correspond to the state energies obtained following proton transfer from DF to each of the ($n - 1$) solvent molecules in the simulation, with the corresponding values of ε allowing energy shifts of the diagonal state energies. To calculate the diagonal elements, we utilized the functional forms and parameters available in the Merck Molecular Mechanics force field,[52] with some important modifications. In particular, the default van der waals parameters of the D atom in DF were changed from their default values to correspond to those of the H in $H_2O$. This was required in order to give the appropriate post-reaction complex stabilization energy discussed below (and shown in Fig 7). The charges on D and F were chosen so as to give a DF molecular dipole moment in agreement with that obtained from gas-phase density functional theory calculations. In addition, we modified the standard MMFF force-field setup to allow for the existence of (1) the F radical, (2) $sp^2$ hybridized radicals of the sort that occur in the nascent $CD_2CN$ co-product, (3) the $CD_2CND^+$ and $CD_3CND^+$ valence states.

The off-diagonal matrix elements $H_{ij}$ are responsible for coupling together diagonal diabatic states $i$ and $j$ in Fig 2. In particular, state 2 couples to the proton transfer state of every solvent molecule, and we assumed that the coupling parameters describing these interactions were identical (i.e., the coupling has the form of $H_{24}$) regardless of the solvent molecule's identity. We further assumed that $H_{24} = H_{23}$. This was motivated by the fact that the proton transfer energies on the cyano end of acetonitrile are largely insensitive to whether the Carbon on the opposite side is $sp^2$ or $sp^3$ hybridized.

In the results described below, we modelled the reactive dynamics of an F radical embedded in 62 solvent molecules, giving a Hamiltonian matrix with dimensions of 64 × 64. The computational cost of these simulations is just over 64 times as large as the cost of running a typical simulation which only involves one state. The decision to utilize a 64-state matrix was determined through consideration of the number of CPU cores which we could reasonably exploit on the architectures available to us, the maximum size of the simulation required so as to quench DF without heating of the bath, and the fact that the our computational resources consisted of 8-core CPU nodes. To reduce the waiting time required to complete any given simulation, we implemented a parallelized dynamics propagation strategy which is schematically illustrated in Fig 3. The propagation algorithm works by instructing each diabatic state to calculate its energy and forces in parallel as a separate MPI CHARMM process. The results for each state are then gathered together to construct the matrix elements for the

Hamiltonian in Eq (1). This matrix is subsequently diagonalized to obtain its eigenvalues and eigenvectors, i.e.:

$$\mathbf{D} = \mathbf{U}^T \mathbf{H} \mathbf{U} \qquad (2)$$

where **D** is a diagonal matrix which contains the eigenvalues, $\lambda$, and **U** is a matrix of eigenvectors. The adiabatic ground state energy is taken as $\lambda_0$, the lowest eigenvalue of **D**, with the corresponding eigenvector $\mathbf{U}_0$ containing the coefficients whose square describes how each diabatic basis state contributes to the state with energy $\lambda_i$. Application of the Hellman-Feynman relation then gives a matrix of Cartesian atomic forces **F**:

$$\mathbf{F} = -\frac{d\mathbf{D}}{d\mathbf{q}} = \mathbf{U}^T \frac{d\mathbf{H}}{d\mathbf{q}} \mathbf{U} \qquad (3)$$

where $\mathbf{F}_0$ is a vector containing those forces which correspond to the lowest eigenvalue. The forces $\mathbf{F}_0$ and the eigenvalue $\lambda_0$ are then dispatched to each MPI process, to overwrite the force and energy data on each process. Each MPI process then propagates forward a single dynamical timestep; the identical forces and energies ensure that each process propagates to an identical geometry. At the new geometry, each process carries out its own energy and force calculations, the results of which are specific to the connectivity of the particular diabatic state. Because force calculations are notoriously the most expensive part of classical MD propagation schemes, this parallelized propagation strategy scales nearly linearly so long as a large enough multi-core architecture is available for use during simulations. The only additional cost is that required to diagonalize the Hamiltonian matrix and calculate the Hellman-Feynmann forces at each time step.

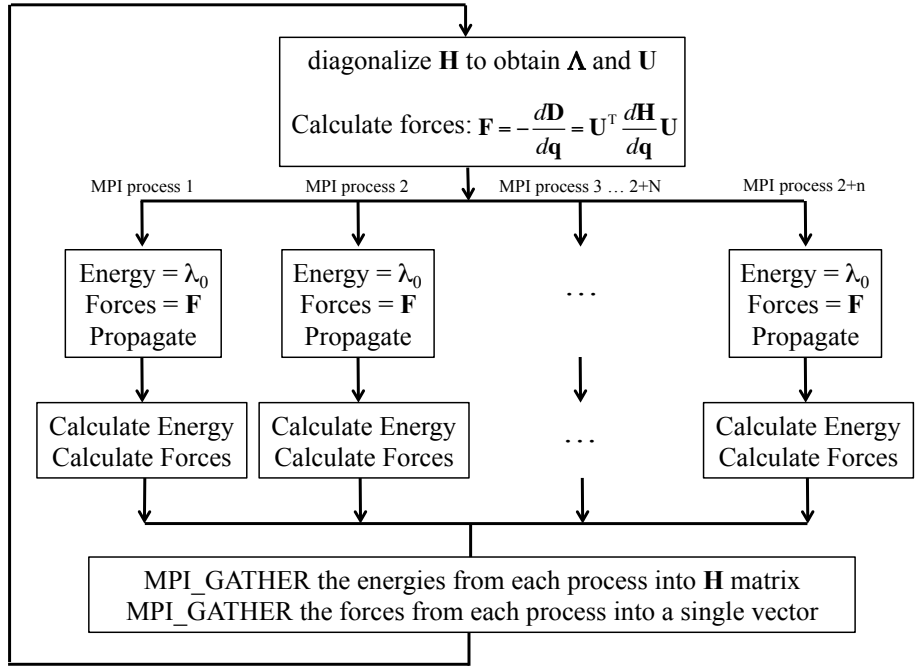

Figure 3: MPI parallelized propagation scheme for efficient molecular dynamics propagation with simulations utilizing large EVB matrices

The trickiest aspect of the EVB method involves finding an appropriate functional form and parameter values for the off-diagonal matrix elements, $H_{ij}$. In our previous work, we modelled these off-diagonal elements using one-dimensional Gaussian functions of interatomic distance. We tested the same approach for the F abstraction energies in this system – i.e., choosing a one-dimensional Gaussian which was a function of the F-H distance. However, functional forms of this type yielded relatively poor fits to the 2d PES data shown in Fig 4. To obtain satisfactory fits, we instead utilized two-dimensional ellipsoid Gaussian functions of the form:

$$H_{12}(r_1, r_2) = A_{12} \exp\left(-(a(r_1 - r_1^0)^2 + 2b(r_1 - r_1^0)(r_2 - r_2^0) + c(r_2 - r_2^0)^2)\right) \tag{4}$$

where $r_1$ is the interatomic D–F distance, $r_2$ is the interatomic C–D distance, $A_{12}$ is the Gaussian amplitude, and $r_0^1$ and $r_0^2$ the respective equilibrium values of $r_1$ and $r_2$. The $a$, $b$, and $c$ parameters are defined as follows:

$$a = \frac{\cos^2\theta}{2\sigma_{r1}^2} + \frac{\sin^2\theta}{2\sigma_{r2}^2}$$

$$b = \frac{\sin 2\theta}{4\sigma_{r1}^2} + \frac{\sin 2\theta}{4\sigma_{r2}^2} \qquad (5)$$

$$c = \frac{\sin^2\theta}{2\sigma_{r1}^2} + \frac{\cos^2\theta}{2\sigma_{r2}^2}$$

where $\theta$ is the Gaussian rotation angle, and $\sigma_{r1}$, $\sigma_{r2}$ are the respective widths in the $r_1$ and $r_2$ directions. While two-dimensional Gaussians were required to obtain satisfactory fits in the abstraction region of the potential, the same was not true for electronic structure points obtained from scans over the post-reaction complex: here we found no advantage of two-dimensional Gaussians over one-dimensional Gaussians – i.e.:

$$H_{24}(r) = A_{24} \exp\left(-\frac{1}{2}([r-r_0]/\sigma)^2\right) \qquad (6)$$

where $r$ in this case is equal to the DF interatomic distance, $A_{24}$ is the Gaussian amplitude, and $\sigma$ is the width parameter.

To determine the values of the Gaussian parameters in Eq (4) – (6), we implemented a Levenburg-Marquardt nonlinear least squares algorithm to fit the EVB model Hamiltonian to the CCSD(T) results. The merit function used to determine goodness of fit was:

$$\chi^2(\text{EVB parameters}) = \sum_{q \in \text{scan points}} \left[\frac{\lambda(\mathbf{q}) - (E_{CCSD(T)}(\mathbf{q}))}{E_{CCSD(T)}(\mathbf{q})}\right]^2 \qquad (7)$$

For the two-dimensional Gaussian in Eq (4) – (5), the EVB parameters included $A_{12}$, $\theta$, $\sigma_{r1}$, $\sigma_{r2}$ $r_0^1$ and $r_0^2$; for the one-dimensional Gaussian in Eq (6), the EVB parameters included $A_{24}$, $r_0$, and $\sigma$. Additional float parameters included $\varepsilon_1$ and $\varepsilon_2$ in Eq (1), which were chosen to give the correct reaction energy. Using this methodology, fits to the Flourine-Hydrogen atom abstraction pathway, obtained by scanning over the C-D and F-D distances, are shown in Figure 4. Fits to the post-reaction CD$_3$CN-DF complex, obtained by scanning over the CN-DF and CND-F distances, are shown in Figure 5. The final set of optimized parameters is given in Table 2. The root mean squared (RMSD) average error between the fitted PES points, and the CCSD(T) points in Fig 4 is 1.05 kcal mol$^{-1}$, and the RMSD for Fig 5 is 2.10 kcal mol$^{-1}$.

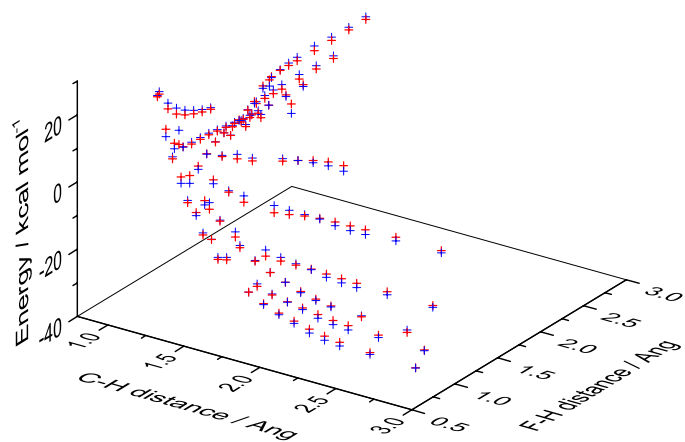

Figure 4: comparison between the CCSD(T) energies (blue) and the optimized MS-EVB model (red) for geometries sampled along the abstraction reaction path

Table 3: best fit parameters obtained from non-linear least squares fitting of the EVB model to the CCSD(T) results

| $H_{12}$ (2d Gaussian) | | $H_{24}$ (1d Gaussian) | |
|---|---|---|---|
| $A_{12}$ | 96.6 | $A_{24}$ | 36.8 |
| $\theta$ | 0.161 | $\sigma$ | 0.306 |
| $\sigma_{r1}$ | 0.311 | $r_0$ | 1.22 |
| $r_0^1$ | 1.61 | - | - |
| $\sigma_{r2}$ | 1.84 | - | - |
| $r_0^2$ | 4.18 | - | - |

Table 4: Frequencies of species used in molecular dynamics force field simulations

| Species | MD force field |
|---|---|
| DF | 3000 |
| $CD_3CN$ | 407, 407, 801, 811, 811, 1026, 1026, 1102, 2101, 2243, 2243, 2271 |
| $CD_2CN$ | 328, 409, 526, 840, 881, 1104, 2169, 2307, 2333 |
| $CD_2CN–DF$ | 34, 35, 183, 320, 337<br>354, 423, 531, 840, 885, 1106, 2169, 2301, 2333<br>2617 |
| $CD_3CN–DF$ | 28, 28, 211, 360, 360<br>423, 423, 811, 811, 811, 1026, 1026, 1104, 2101, 2243, 2243, 2270<br>2652 |

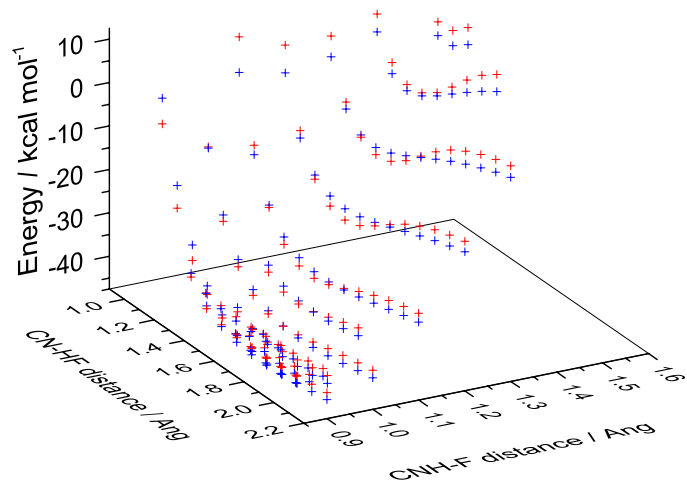

Figure 5: comparison between the CCSD(T) energies (blue) and the optimized MS-EVB model (red) for geometries sampled along in the vicinity of the CD$_3$CN-DF complex

Having carried out the fits using the data in Figs 4 and 5, we used the Eq (1) model to calculate several other parts of the global potential energy surface. First, we calculated the optimized one dimensional minimum energy abstraction path. The results are shown in Fig 6, and reveal a forward reaction barrier of 2.5 kcal mol$^{-1}$, in good agreement with the transition state barrier height predicted by our CCSD(T) calculations. Second, we carried out an optimized one dimensional scan in the post-reaction region of the CD$_3$CN-DF potential. The results are shown in Fig 7, and reveal a post-reaction complex with a stabilization energy of just over 7 kcal mol$^{-1}$, which again agrees well with the results obtained from the CCSD(T) calculations. Finally, we used the EVB model to carry out a 2d scan over the C-D and F-D distances in the post reaction CD$_2$CN-DF complex. Accurately modelling these results requires coupling states 2 and 3 in Fig 2 (i.e., coupling element $H_{23}$ in Eq (1)). Rather than introduce new parameters to describe this coupling element, we assumed that it was identical to $H_{24}$ in both its functional form (i.e., representing it as a one-dimensional Gaussian that depends on the interatomic DF distance) and its parameter values. The results obtained using this approximation are shown in Fig 8, along with a comparison to the MS-EVB model energies obtained for the same scan points in the CD$_3$CN-DF complex. The good agreement in energy between points calculated in the region of both CD$_3$CN-DF and CD$_2$CN-DF is reassuring, and suggests some transferability in the functional form of the coupling element as well as the values of the optimized parameters.

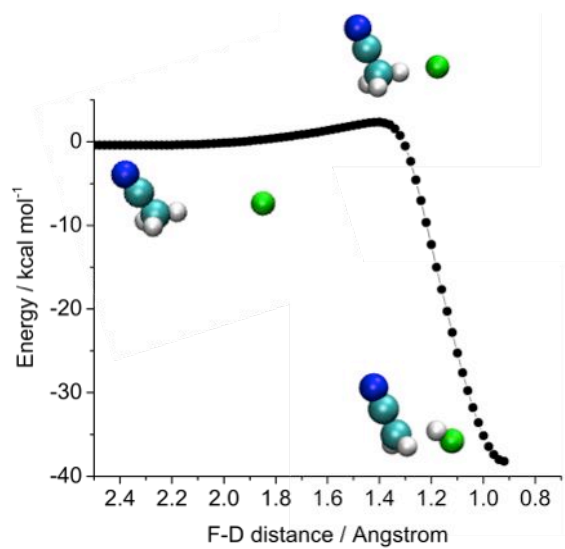

Figure 6: MEP for the abstraction pathway, F + CD$_3$CN → DF + CD$_2$CN

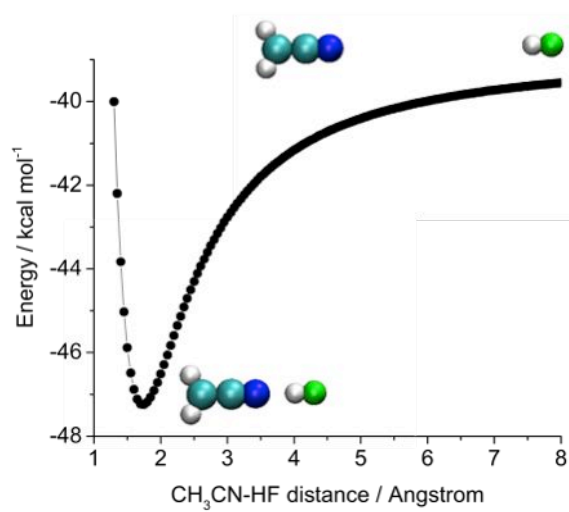

Figure 7: relaxed scan along the CD$_3$CN-DF distance in the solute/solvent complex

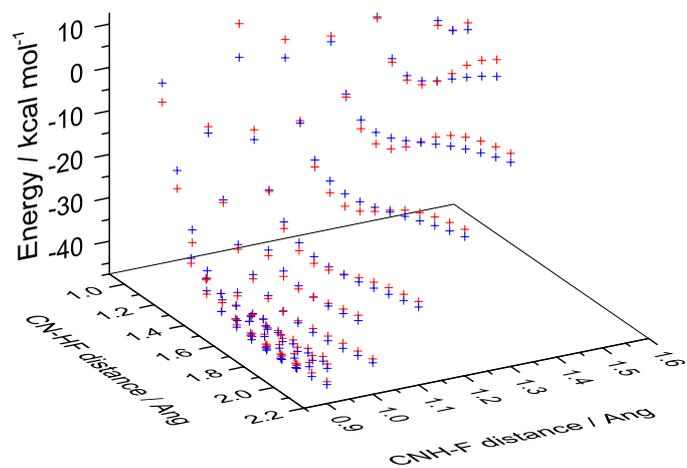

Figure 8: comparison of results obtained using the optimized MS-EVB model for the CD$_3$CN-DF complex with MS-EVB results obtained for the CD$_2$CN-DF complex

## Dynamics Simulations

*Methods & Software*

All of the dynamics work described in this paper was carried out using a locally modified version of the CHARMM software suite, to which we recently added general routines and an associated input structure that allows the user to specify EVB Hamiltonians with matrix elements which are linear combinations of 2d Gaussians, 1d Gaussians, and/or constants. Fitting was carried out using a script to interface with CHARMM with the Levenburg-Marquardt algorithm implementation available within the scientific python (SciPy) library. The diagonal elements of the Hamiltonian matrix were calculated using the Merck Molecular force-field (MMFF) in CHARMM,[52] with the modifications discussed above.

All dynamics simulations began with NVT equilibration runs that used the leapfrog Verlet integration scheme. These were followed by subsequent NVE trajectories using a velocity Verlet propagation algorithm, resulting in initial conditions that were sampled using classical mechanics and the full EVB Hamiltonian. The discussion below makes reference to several different types of simulations, in which we: (1) simulated gas-phase DF as a Morse oscillator with different initial energy contents, (2) simulated the DF relaxation dynamics, (3) simulated the equilibrium dynamics, and (4) simulated the coupled reaction/relaxation dynamics. All simulations involving solute and solvent were carried out in a periodic cubic box with edge lengths of 17.8 Å.

To investigate DF relaxation dynamics in bulk solvent, we carried out 100 separate simulations with an initial geometry wherein DF was solvated within a periodic box including 61 $CD_3CN$ molecules, and one $CD_2CN$ molecule, in line with the room temperature experimental density of acetonitrile (0.79 g/mL at room temperature). Equilibration runs of 100ps (0.5 fs timestep) with a dissipative Langevin thermostat (friction coefficient of 10 $ps^{-1}$, and a heat bath of 300K) were used to generate an ensemble of initial coordinates and velocities. These coordinates and velocities were used as starting points for subsequent NVE trajectories, with a duration of 10ps (0.1 fs timestep). Before launching the NVE trajectories, the velocity of the D atom in DF was given a non-equilibrium 'kick' of ~35 kcal $mol^{-1}$ in the direction of its bonded Fluorine neighbor using a local-mode approach of the sort described previously.[53] This quantity of energy roughly corresponds to a DF vibrational quantum number of 4, and essentially corresponds to the limit that all of the excess energy available following (R1a) is deposited in the product DF. To examine the properties of DF solvated by $CD_3CN$ solvent in a regime close to equilibrium, we carried out 50 separate simulations with an initial geometry setup identical to that utilized in the relaxation simulations. NVT trajectories were used to generate starting points for subsequent NVE propagation in a manner identical to the relaxation trajectories, with the exception that no initial non-equilibrium kick to the H atom was implemented.

For the reaction dynamics, we initialized the simulations with an F radical embedded in 62 $CD_3CN$ solvent molecules, and carried out 200 separate reactive trajectory simulations using equilibration runs of 100ps (0.5 fs timestep) with a dissipative Langevin thermostat (friction coefficient of 10 $ps^{-1}$, and a heat bath of 300K). The ensemble of initial coordinates and velocities generated in these trajectories were used to launch subsequent NVE trajectories, with a duration of 20ps (0.1 fs timestep). To guarantee that every one of the NVE trajectories resulted in a reaction and thereby improve the statistics of the analyses carried out below, we exploited the recently developed BXD class of algorithms.[54-56] BXD is a formally exact extension of transition state theory (TST), which is particularly well suited to accelerating reactive events in studies such as this. So long as the distance between the constraints and the transition state is larger than the system's characteristic dynamical decorrelation length, then a BXD-accelerated simulation conserves energy and gives meaningful statistics following transition state passage. In non-equilibrium studies like that outlined herein, this is a considerable advantage of BXD compared to other biasing methods – e.g., umbrella sampling, where the biasing has the consequence that the dynamical results are no longer meaningful in non-equilibrium regimes.[54,55] During the equilibration runs, we specified BXD constraints to ensure that the distance between the Fluorine radical and the reactive H had an upper bound of 1.5 Å and a lower bound of 1.8 Å, which is well on the reactant side of the abstraction TS. These constraints prevented the reactants diffusing away from one another while still preserving interactions of the reactants with the neighboring solvent molecules. In the NVE runs, the lower bound BXD constraint was relaxed, accelerating the rate of transition state passage and resulting in every trajectory undergoing an abstraction event, usually within 0.5 ps of the first timestep.

Upon their completion, all of the trajectories described above were examined to ensure that they satisfied energy conservation to within better than 1% of the total kinetic and potential energy. It is not uncommon that dynamics simulations carried out utilizing multi-state EVB methods fail to conserve energy, owing to an incomplete basis set of valence states in Eq (1). Because our simulations included all possible couplings for a specified valence state (i.e., the Hamiltonian matrix included the interaction of DF with every possible solvent molecule), they were not subject to this source of error. For example, every equilibrium trajectory conserved energy within the specified 1% acceptability threshold. The reaction and relaxation dynamics, on the other hand, were subject to energy conservation problems; however, it was not the MS-EVB model that was the source of these failures. Rather, they arose because of the large amount of energy (~35 kcal mol$^{-1}$) initially localized in the DF stretching motion. As a result, on the order of 20% of the reaction and relaxation dynamics failed to properly conserve energy. These were excluded from the analyses described below.

Time dependent energies of DF were determined using the strategy outlined in previous work.[36, 37] Briefly, DF's time-dependent Cartesian velocities, $\dot{\mathbf{q}}(t)$, (obtained from non-equilibrium NVE trajectories) were projected into the translational, rotational, and vibrational normal modes of DF in its center-of-mass frame equilibrium geometry, $\mathbf{q}_{eq}$, as follows:

$$\dot{\mathbf{Q}}(t) = \mathbf{L}^{-1} \dot{\mathbf{q}}(t) \tag{8}$$

where $\dot{\mathbf{Q}}(t)$ is a vector of normal mode displacements and velocities, and $\mathbf{L}$ is a 3N × 3N matrix obtained from diagonalizing the mass-weighted Hessian of DF (with column vectors corresponding to the Cartesian displacements of DF's 3 translations, 2 rotations, and single vibration). The kinetic energy of the $\ell$'th normal mode, $\mathbf{T}^{\ell}(t)$, is determined as

$$\mathbf{T}^{\ell}(t) = \sum_{i\alpha} \frac{m_i}{2} [\mathbf{L}^{\ell}_{i\alpha} \dot{\mathbf{Q}}^{\ell}(t)]^2 \tag{9}$$

where $m$ is the atomic mass, $i$ runs over the atom indices, and $\alpha$ runs over the Cartesian $x$, $y$, $z$ directions. In this notation, the column vectors in $\mathbf{L}$ have been transformed from mass-weighted to Cartesian space, and subsequently normalized using the appropriate normalization constant, $N_{\ell}$. Unlike Cartesian velocities, the kinetic energy is diagonal in the normal mode displacements. The virial theorem specifies that, on average, the total energy is equipartitioned between kinetic and potential contributions, so that the average total energy in some mode over a particular time window $\tau$, may be calculated as:

$$\langle \mathbf{E}^{\ell}(t) \rangle \approx 2 \langle \mathbf{T}^{\ell}(t) \rangle_{\tau} \tag{10}$$

where the angled brackets indicate averages. All reported values of the DF stretching energy obtained in this work used Eq (10). So long as $\tau$ spans several vibrational periods of the stretching mode, then Eq (10) may be expected to give reasonably accurate results.[57] For analysis of the DF stretch, the averaging was carried out with $\tau = 250$ fs.

The spectra reported in the analysis that follows were obtained from the well-know relationship that links a power spectrum to the Fourier transform of some dynamical observable $C(t)$:[58, 59]

$$I(\omega) = \frac{1}{2\pi} \int_{-\infty}^{+\infty} C(t) \exp(-i\omega t) dt \qquad (11)$$

Eq (11) is often cast in an alternative form that permits one to utilize faster Fourier algorithms to obtain power spectra from dynamical observables which have a finite time duration, $2T$:[58-60]

$$I(\omega) = \frac{1}{2\pi} \lim_{T \to \infty} \frac{1}{2T} \left\langle \left| \int_0^{2T} C(t) \exp(-i\omega t) dt \right|^2 \right\rangle \qquad (12)$$

where the angled brackets indicate an average over trajectories launched with different sets of initial conditions. In this work, $C(t)$ was taken to be the velocity autocorrelation function, i.e., $<v(0) \cdot v(t)>$ where $v$ is a vector containing all the velocities of a relevant set of atoms. The spectral results reported herein utilize a sampling frequency of 1 fs (i.e., sampling every 10 time steps), which according to the Nyquist theorem, allow us to detect periodic motion with frequencies of ~16,000 cm$^{-1}$ or less. The spectral resolution of Eq (12) depends on how long of a time window, $2T$, is spanned by the correlation function (longer time spans allow increasingly fine resolution). All time-dependent spectra were calculated from correlation functions with a length of 1.024 ps (i.e., $2T = 1024$ fs). In those plots which include a sequence of time-dependent spectra, obtained from a sequence of correlation functions, individual spectra are indexed by the midpoint of the time window spanned by $C(t)$ (e.g., the spectra obtained from the correlation function spanning 0 to 1.024 ps is referred to as the '0.512 ps' spectra).

In addition to time-dependent spectra, we also report time dependent radial distribution functions (RDFs) to analyze transient changes in the DF solvent environment. All radial distribution functions are calculated in a manner very similar to the time dependent spectra. The radial distribution functions presented below were obtained by averaging together the RDFs obtained from separate trajectories. Individual RDFs for a given trajectory were obtained by calculating interatomic distances every femtosecond (i.e., every 10 timesteps), placing them in data arrays of length 1024, and subsequent histogramming of the 1024 member data arrays. All reported RDFs were constructed using histogram bins of 0.05 Angstrom. Normalization of each RDF was carried out following averaging. Data arrays with a length of 1024 were chosen to maintain consistency with the analyses carried out to construct time-dependent vibrational spectra. To aid interpretation of the transient DF spectra obtained using Eq (12), we utilized a nonlinear least squares minimization procedure to fit the raw spectra, $I(\omega)$, to a sum of Gaussian functions as follows:

$$I(\omega) = \sum_{i=1,2} A_i \exp\left[-\frac{(\omega - \omega_i^0)^2}{2\sigma_i^2}\right] \quad (13)$$

where $A_i$ is the amplitude of Gaussian function $i$, $\sigma_i$ is its corresponding width, and $\omega_i^0$ is the position of its center. At most, we carried out fitting using two Gaussian functions ($i = 1, 2$), which we found adequately captured the DF features obtained using Eq (12). In general, a single Gaussian adequately captures the transient behavior of the dominant DF stretching peak which is the emphasis of this work. The utility of the two-Gaussian approach is its ability to capture a small amplitude spectral feature to the blue of the main DF stretching feature, which we observed in a number of our dynamics simulations. In general, this smaller spectral feature corresponds to DF which is not engaged in a strongly bound solvent H-bond complexes. While this is an interesting observation, the qualitative conclusions derived from the two-Gaussian fits are more or less identical to those from the one-Gaussian fits. In the text that follows, we refer to the results obtained from the one-Gaussian fits. The SI includes results obtained from two-Gaussian fitting.

***Gas Phase Dynamics of DF Morse Oscillator***

The time-dependent DF spectra obtained from reaction dynamics simulations of (R1) are the result of a complex set of competing effects, the first of which concerns the anharmonicity of the DF stretching mode. Indeed, it is the anharmonicity in the DF stretch which permits spectral identification of transient vibrational excited states. From the quantum mechanical perspective, allowed transitions between adjacent vibrational states have an energy which decreases linearly with increasing vibrational quantum state.[61] Consequently, adjacent transitions at higher vibrational states lie to the red of transitions at lower lying vibrational quantum states. From the classical perspective, where vibrational eigenstates are not quantized, the vibrational energy content in the Morse oscillator is on a continuum; however, the vibrational frequency of the oscillator red shifts as a function of energy owing to the increasing importance of large amplitude anharmonic motions. Because the goal of this work is to learn about transient highly vibrational excited states of DF formed through chemical reactions in condensed phase systems, our simulation approach necessarily utilizes a classical approach, especially given the computational expense incurred with our 64-state MS-EVB potential energy function.

Classical approaches are unable to capture the quantized transitions between vibrational eigenstates which occur in quantum mechanical approaches, with the net result that spectra obtained from classical simulations lack the structure seen in quantum mechanical approaches. However, these fine structures are often washed out in condensed phase systems, and previous work has shown that the classical approach to calculating diatomic vibrational spectra can often provide line shapes which agree very well with those obtained in both quantum mechanical calculations and experimental observations.[59, 62] For systems in their ground vibrational state, there is a well-known systematic error in the peak locations calculated from classical spectra compared to their quantum mechanical counterpart spectra, which may be easily corrected by a simple energy shift formula.[59] Anharmonicity is the principle source of this error: a classical oscillator has an energy on the order of $k_BT$, meaning that it is confined to a largely harmonic region at the bottom of the Morse potential, whereas a quantum oscillator has a minimum energy which corresponds to its $v = 0$ zero point (i.e., several times $kT$). Observed spectral features primarily arise from transitions between $v = 0$ and $v = 1$, meaning that the quantum oscillator therefore samples larger regions of the anharmonic phase space compared to the classical system. The extent of disagreement between the classical and quantum mechanical approaches is therefore most dramatic at low energies – i.e., close to the thermal regime. In this work, where the DF is produced from chemical reaction with substantial vibrational excitation ($v \sim 2 - 3$), then: (1) the initial vibrational energy content in the nascent DF will be very similar whether it is treated classically or quantum mechanically; and (2) detailed balance requires that downward transitions will dominate compared to upward transitions. Consequently, in condensed phase regimes with oscillators that have a high initial energy, and so long as the ensemble-averaged downward transition rates are approximately equal in the classical and quantum simulations, it is reasonable to suppose that the classical and quantum mechanical systems will explore similar regions of the anharmonic phase space, giving a smaller deviation between classical and quantum mechanical spectra.

A common theoretical approach for calculating vibrational energy relaxation (VER) rate coefficients from state $i$ to $j$ is to split the simulation into 'system' and 'bath' components. The coupling, which allows energy to flow between the system and bath, is the Fourier transform of the quantum correlation function of the $ij$ system matrix coupling element.[8, 63] However, accurate calculation of the quantum time correlation is extremely difficult for all but the smallest systems. Hence, a more common approach is to replace the quantum time correlation function with a classical time correlation function, with the subsequent application of a quantum correction factor (QCF). For ground state vibrational transitions, a number of formulas have been derived which provide QCFs to classical vibrational energy relaxation results.[6, 8, 63, 64] For low energy $v = 1 \leftarrow 0$ transitions, the quantum correction factors are often small, (on the order of 2 – 3) so long as one takes care to use reasonably accurate force fields.[6, 63] VER is notoriously sensitive to the system bath coupling. In condensed phase systems, it is therefore usually difficult to determine whether discrepancies between calculated VER and experimental VER arise from the QCF, and not from errors in the potential.[8] For higher energy transitions, the form of the QCF remains subject to substantial uncertainty.

These uncertainties in the form that the QCF should take for higher lying transitions, coupled with the added complexity of as a result of the fact that the transient high energy states which we are investigating arise from a chemical reaction event, led us to utilize a purely classical approach. QCFs act to increase the rate of VER compared to the classical result; therefore, our results provide a lower limit on the rate at which DF relaxes in $CD_3CN$ solvent. As shown in what follows, our results agree very well with the available experimental data, but the complexities of the system under investigation, combined with the experimental errors, making it difficult to quantitatively assess the relative importance of small QCFs.

To represent DF as a Morse oscillator, we added a subroutine to CHARMM which allows the user to select any given harmonic bond and assign it the standard Morse functional form:

$$V(r) = D_e(1 - \exp(-a(r - r_e)))^2$$
$$a = \sqrt{k_e / D_e}$$
(14)

where $r$ is the bond distance, $D_e$ is the bond dissociation energy, $r_e$ is the equilibrium bond distance, and the relationship between $a$ and $k_e$ can be seen through a Taylor series expansion of Eq (14). The value of $k_e$ was set to 9.657 millidyne Å$^{-1}$. With this value of $k_e$, the frequency at the bottom of the Morse well, $v_0$ (obtained by diagonalization of the diatomic Hessian) for HF and DF are 4138 cm$^{-1}$ and 3000 cm$^{-1}$, respectively, in close agreement with the corresponding experimental gas-phase values of 4138 cm$^{-1}$ and 2998 cm$^{-1}$ (NIST webook). The value of $D_e$, determined from CCSD(T)-F12b/aug-ccpVTZ calculations, was set at 141.28 kcal mol$^{-1}$. Figure 9 shows a comparison between a range of DF geometries calculated at the CCSD(T)-F12b/aug-ccpVTZ level, and the corresponding analytic form for the Morse potential (Eq 14) implemented within CHARMM for use in these simulations.

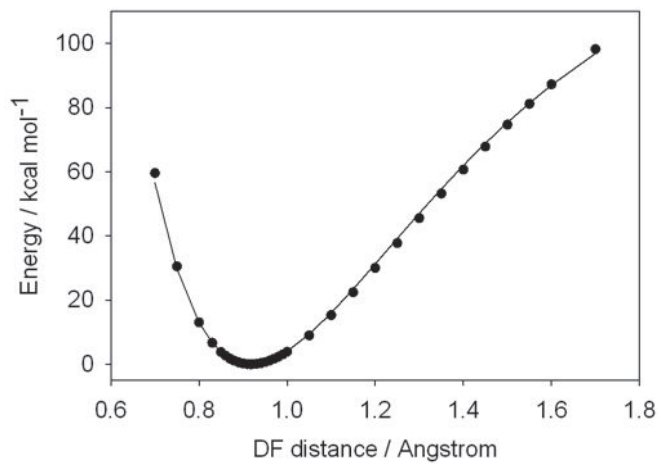

Figure 9: comparison between CCSD(T)-F12b/aug-ccpVTZ and the analytical Morse function used in the dynamics simulations

To characterize anharmonic shifts which arise solely from the Fig 9 Morse potential which we used to describe the DF, and also to verify that our implementation of Eq (14) gives accurate DF vibrational spectra, we carried out gas-phase DF simulations, with initial velocities selected to give different initial DF vibrational energies (always with zero rotational energy). Analytical solution of the classical equations of motion for a Morse oscillator predicts an energy dependent red shift in the vibrational frequency with increasing energy, which goes as:[61, 65]

$$v(E) = v_0 \sqrt{\frac{D_e - E}{D_e}} \qquad (15)$$

where $E$ is the total kinetic + potential energy of the Morse oscillator relative to the potential energy minimum, and $v_0$ is the harmonic frequency (3000 cm$^{-1}$ for our DF model). The results obtained from Eq (15) are given in Table 3 at a range of different Morse energies, up to a maximum of 36.9 kcal mol$^{-1}$, which corresponds to essentially all of the excess energy of (R1a) going into DF vibrational energy. Table 3 shows that, within errors which result from discrete sampling of the DF vibrational spectra over a finite time interval, the agreement between the Eq (15) energy dependent analytical frequencies and those obtained using the spectral approach of Eq (12) agree very well. At a vibrational energy of 36.9 kcal mol$^{-1}$ (which corresponds to essentially all of the excess energy of (R1a) going into DF vibrational energy), the DF frequency from spectral analysis is 2578 cm$^{-1}$, a value which is red shifted 422 cm$^{-1}$ with respect to $v_0$.

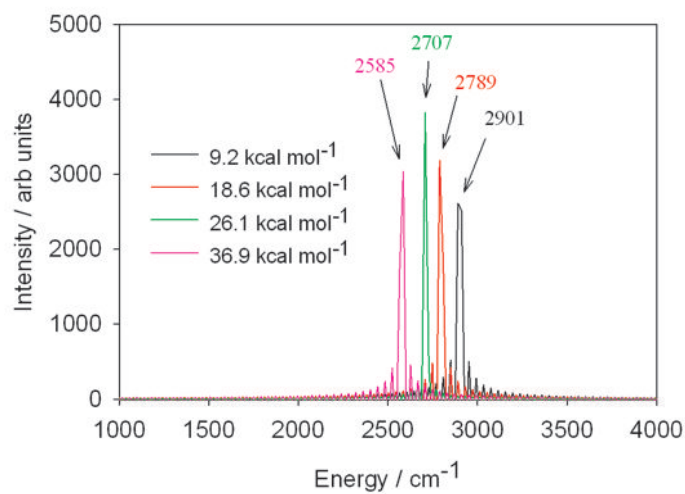

Figure 10: vibrational spectra obtained from gas phase simulation of isolated DF, with different initial vibrational energies, and analyzed using the spectral analysis approach of Eq (12).

Table 5: Comparison of vibrational frequencies obtained for a classical Morse oscillator at a range of different energies, using the spectral analysis outlined in Eq (12) and the analytical solution of Eq (14)

| Total Energy | Vibrational Frequency from Eq (15) | Vibrational Frequency from Eq (12) Power Spectrum |
|---|---|---|
| 9.2 kcal mol$^{-1}$ | 2900 cm$^{-1}$ | 2901 ± 10 cm$^{-1}$ |
| 18.6 kcal mol$^{-1}$ | 2796 cm$^{-1}$ | 2789 ± 10 cm$^{-1}$ |
| 26.1 kcal mol$^{-1}$ | 2708 cm$^{-1}$ | 2707 ± 10 cm$^{-1}$ |
| 36.9 kcal mol$^{-1}$ | 2578 cm$^{-1}$ | 2585 ± 10 cm$^{-1}$ |

*Vibrational Relaxation Dynamics*

Having examined the spectra of the isolated DF Morse oscillator, we next investigated the DF relaxation dynamics in CD$_3$CN solvent, without any reaction. Initially, we tested the simplest possible potential energy surface, with a 2 × 2 EVB matrix only including the $V_1 + \varepsilon_1$, $V_2 + \varepsilon_2$, and $H_{12}$ terms in Eq (1). While this potential was satisfactory insofar as it gave a post-reaction complex with an energy of ~8 kcal mol$^{-1}$ akin to that shown in Fig 7, it nevertheless gave extremely slow energy relaxation rates, as discussed below. Since early work carried out by Forster, several experimental and theoretical studies have shown that energy transfer depends strongly on (1) the overlap between the solute and solvent spectra, and (2) the solvent/solute coupling. In the case of DF solute embedded in CD$_3$CN solvent, spectral decomposition of the simulation data utilizing a 2 × 2 EVB matrix gives a solvent band between ~2000 – 2340 cm$^{-1}$ (with well-defined peaks near 2101 cm$^{-1}$ and 2247 cm$^{-1}$, see Fig 14), corresponding to the CN stretching frequency. With larger separation between the DF frequency and the nearest solvent bands, the spectral overlap is weaker between donor DF vibrations and acceptor CN vibrations, resulting in slower energy transfer. To probe the dependence of DF energy relaxation on the solute/solvent spectral structure, we carried out a sensitivity analysis of energy transfer as a function of the DF force constant. The results are shown below in Figure 11, and were obtained by averaging over different sets of 10 trajectories with ~35 kcal mol$^{-1}$ initially localized in the DF stretch, and each with a different value of the DF force constant. Inspection of Fig 11 indeed shows that energy relaxation from DF into the CD$_3$CN solvent bath increases as the DF frequency approaches that of the CD$_3$CN spectral bands. All of the DF relaxation curves in Fig 11 may be well fit using a single exponential function. Using a DF force constant which reproduces the experimental gas phase vibrational frequency of 3000 cm$^{-1}$ gives extremely slow energy relaxation, with a time constant on the order of 764 ps$^{-1}$ (averaged over 100 simulations). Systematically decreasing the DF force constant increases the DF energy relaxation rate, with the maximum relaxation rate having a time constant of ~33.4 ps$^{-1}$ occurring at a DF frequency of 2400 cm$^{-1}$ (obtained from averaging over 10 simulations). Further decreasing the DF force constant diminishes its vibrational spectral overlap with the CD$_3$CN bands, and a slower rate of energy transfer. For example, a DF frequency of 2300 cm$^{-1}$ gives a relaxation rate with a time constant of ~76.9 ps$^{-1}$.

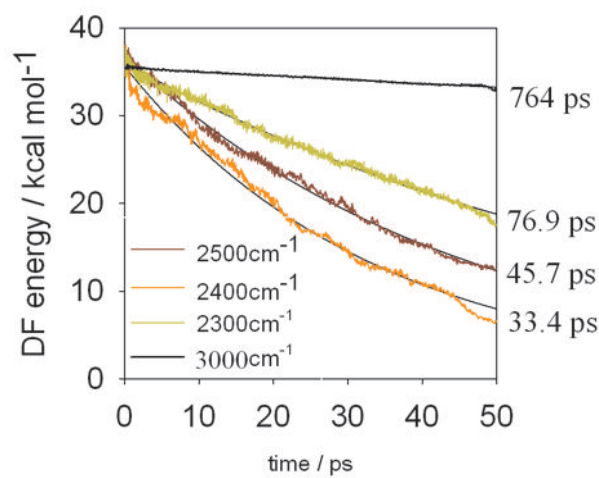

Figure 11: Sensitivity of DF relaxation to the DF force constant. The plots show the relaxation time profiles of DF vibrational energy into the CD$_3$CN solvent, for force constants which give $v_0$ values of 3000 cm$^{-1}$, 2500 cm$^{-1}$, 2400 cm$^{-1}$, and 2300 cm$^{-1}$. In all tests, ~35 kcal mol$^{-1}$ vibrational energy was initially localized in the DF stretch. The numbers to the right of the plot show the time constants obtained by fitting each curve to a single exponential decay

The simulations carried out to obtain Fig 11 treated solvent/solute coupling using a 2 × 2 EVB matrix – i.e., instead of including solvent/solute coupling within the EVB matrix, it was calculated using standard non-bonded terms in the MMFF force field, which include both van der waals and electrostatic interactions. This coupling is rather weak, and independent of the DF bond distance. The Fig 11 results suffer from the following shortcomings: (1) the DF relaxation rates are substantially slower than the timescales suggested by the experimental results,[66] which are on the order of a few picoseconds; and (2) adjustment of the force constant to increase the energy relaxation rate is ultimately an unsatisfactory approach because it biases the DF frequency in a fashion that cannot accurately recover the gas-phase experimental vibrational frequency, and substantially underestimates the solvatochromatic shift. Indeed, it was these shortcomings that led us to investigate how the energetics of the $CD_3CN\cdots DF$ post-reaction complex depended on both the DF stretch and the CN stretch, the results of which were presented above and shown in Fig 5.

Fig 12 offers an interesting comparison to Fig 5. It shows energies obtained utilizing the standard MMFF approach in the region of the post-reaction complex, with no additional coupling beyond van der waals and electrostatic terms (geometries used to construct Figs 5 and 12 are identical, obtained from rigid scans over both the H–F and the N$\cdots$H distance, and a DF force constant chosen to give a gas phase vibrational frequency of 3000 cm$^{-1}$). For comparison, Fig 12 also shows CCSD(T) energies at the corresponding geometries. At low energies, for DF distances close the equilibrium value of 0.92 Å (i.e., in the vicinity of the $CD_3CN\cdots DF$ complex minimum energy), there is reasonable agreement between the MMFF and CCSD(T) energies. In regimes with elongated H–F distances and decreased N$\cdots$H distances, the agreement is substantially worse, with the MMFF force field substantially underpredicting the amount of coupling, and energies which are consequently far too large. In the dynamics simulations, large DF interatomic distances are accessible when the DF has significant vibrational excitation. Inspection of Figure 12 thus offers some qualitative insight into Fig 11, and in particular why the MMFF force-field underpredicts the rate at which DF vibrational energy flows into the solvent: the standard MMFF approach underestimates the degree of solvent/solute coupling, leading to energy transfer which is several orders of magnitude slower than what it would be with a better description. This coupling provides insight into why the energy transfer rates obtained using a 2 x 2 Hamiltonian are orders of magnitude too small.

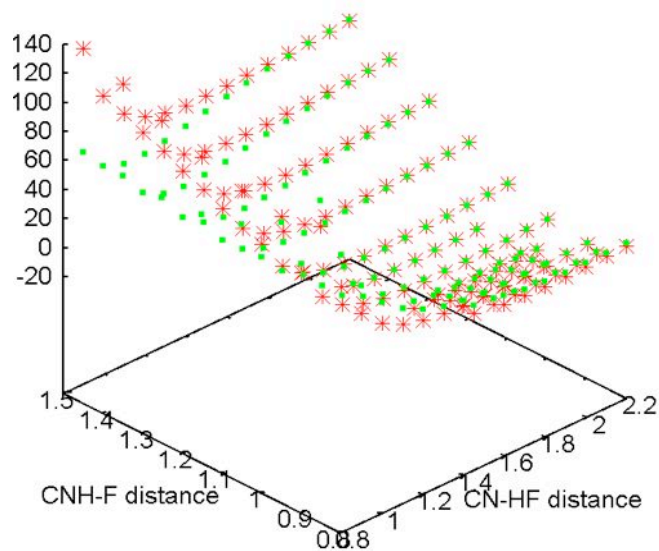

Figure 12: Comparison between the CCSD(T) (green) and MMFF energies (red) in the post-reaction CD$_3$CN⋯DF complex, with scan points obtained as a function of both the H–F and the N⋯H distance. The MMFF energies shown in this graph include no additional CD$_3$CN and DF coupling beyond the van der waals and electrostatic terms available in the standard MMFF methodology. Energies on the z-axis are kcal mol$^{-1}$.

Utilizing a 64 × 64 EVB matrix of the sort outlined in Eq (1) allows us to include the solvent/solute coupling that is missing in the standard MMFF approach. The surface obtained by inclusion of this coupling is shown in Fig 5, and the resultant DF relaxation profile is shown in Fig 13 (~35 kcal mol$^{-1}$ initial excitation energy, similar to the Fig 11 initial conditions). The results in Fig 13 (obtained using a DF force constant chosen to reproduce the gas-phase frequency of 3000 cm$^{-1}$) show substantially faster vibrational energy relaxation of DF following its initial excitation than the simulations results shown in Fig 11. Indeed, within the first few vibrational periods of the initially excited DF, 5 – 6 kcal mol$^{-1}$ of its initial vibrational energy is rapidly transferred to solvent, which accounts for why Fig 11 and Fig 13 appear to have different energies at time zero. Whereas the Fig 11 results were well-fit using a single exponential function, this is not the case for the results in Fig 13. The DF vibrational energy relaxation profile in Fig 13 shows two distinct relaxation regimes – fast relaxation at short times with a time constant of ~0.34 ps, and a long-time relaxation rate which slower by a factor of ~10, with a relaxation time constant of ~5.0 ps. Inclusion of the solvent/solute coupling has a dramatic effect on the DF relaxation rate, giving relaxation timescales which are 2 – 3 orders of magnitude faster than those obtained for the comparable curve in Fig 11. Accurate treatment of the coupling also reproduces the solvatochromatic shift observed between gas-phase DF and DF embedded in CD$_3$CN solvent. Fig 14 shows the equilibrium vibrational spectrum of DF along with that of the CD$_3$CN solvent, both of which were obtained from long equilibrium simulations of DF in CD$_3$CN, with no initial excitation beyond that predicted by the thermal sampling methods described above. Within our simulations, Fig 14 shows that the peak corresponding to the DF stretch (in CD$_3$CN solvent) occurs at 2540 ± 30 cm$^{-1}$, in reasonable agreement with results obtained from experimental IR spectroscopy,[66-68] which indicate that the DF stretch occurs at ~2580 cm$^{-1}$. This corresponds to a solvatochromatic shift of over 400 cm$^{-1}$ compared to the DF gas phase experimental spectra, in good agreement with experimental results.

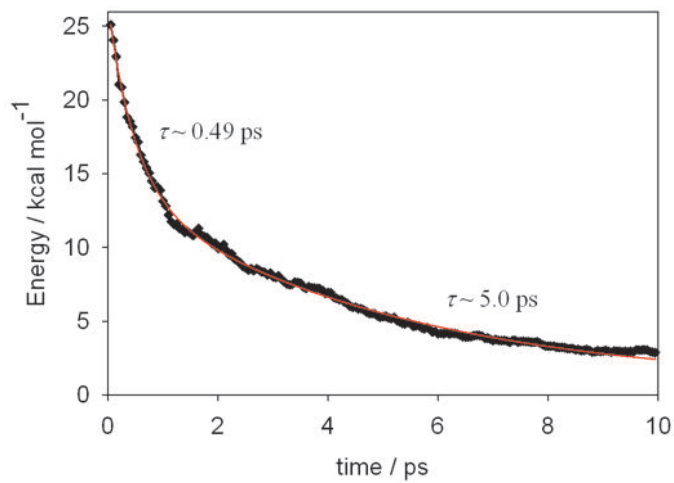

Figure 13: time-profile for vibrational relaxation of DF solute in CD$_3$CN solvent, utilizing the previously discussed 64 × 64 EVB matrix, which includes solvent/solute coupling beyond that included within the standard MMFF approach. Biexponential fits to the relaxation profile clearly shown two distinct regimes – fast relaxation at short times, and slower relaxation at long times

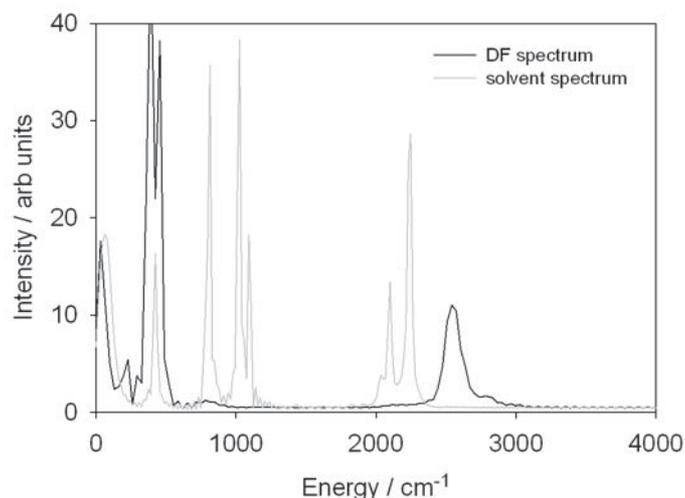

Figure 14: Equilibrium spectra of DF embedded in CD3CN solvent, overlayed with a spectrum of neat CD3CN solvent. Note that the DF spectral data has been arbitrarily scaled so as to clarify its spectral features; otherwise it is dwarfed by the relative magnitudes of the solvent peaks. The DF spectral peak may be well fit with a single Gaussian centered at 2540 cm$^{-1}$.

The multiple timescales observed in the DF relaxation profile (Fig 13) can be partly rationalized through inspection of the time dependent spectra obtained following DF excitation, shown in Fig 15. Black lines show the DF vibrational spectra and grey lines show the equilibrium solvent spectrum. Red lines show fits to the DF stretching feature using a single Gaussian function, while green lines show fits carried out using the sum of two Gaussian functions. The time-dependent results of this fitting procedure are shown in Fig 16. The SI contains additional snapshots of the time dependent spectra, along with plots showing the parameters returned from the two-Gaussian fits at every snapshot. At times less than a picosecond, DF undergoes large amplitude vibrational motion in the immediate aftermath of excitation, resulting in a transient spectrum with a band center of ~2211 cm$^{-1}$ (which is red shifted by over 300 cm$^{-1}$ from its equilibrium solvent position) and a width of ~293 cm$^{-1}$. This results in a very strong overlap with the solvent spectrum peaks at 2101 and 2247 cm$^{-1}$. Combined with the strong coupling that arises from large amplitude DF motion, this gives extremely fast energy relaxation at short times. As time goes on and the DF cools, Figs 15 and 16 show that the DF band sharpens as it blue shifts toward its equilibrium position. This decreases both the extent of solvent/solute spectral overlap, and the probability of large amplitude DF motions that strongly couple to solvent molecules. The combination of these two effects combine to give a time dependent decrease in the energy relaxation rate from the DF solute to the CD$_3$CN solvent, although we cannot rule out the possibility that other mechanisms may also be involved.[69, 70]

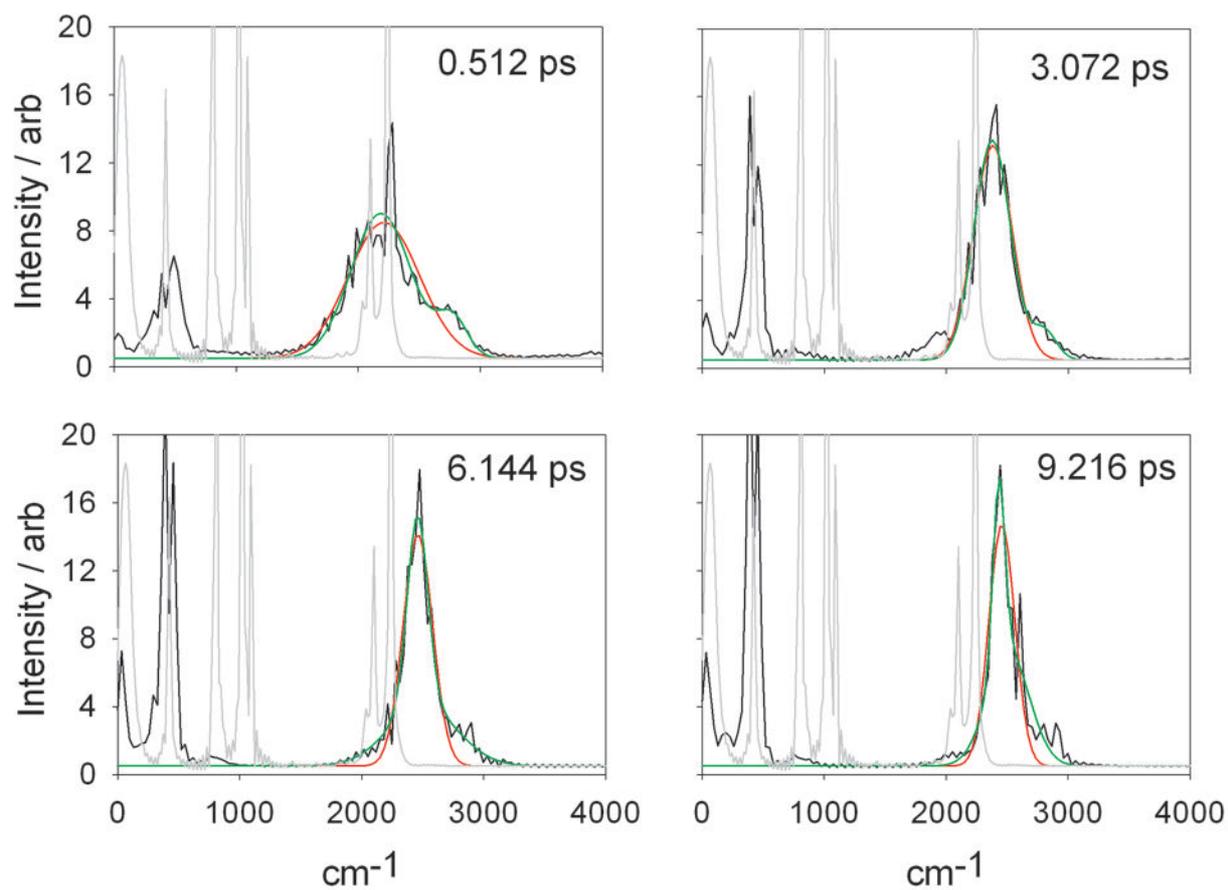

Figure 15: time dependent relaxation spectra of DF following a non-equilibrium 'kick'. The different time slices were obtained by averaging over 1.024 ps time windows to obtain the CD$_3$CN vibrational spectrum (grey) and the DF vibrational spectrum (black). The red and black lines are fits to the DF vibrational spectrum, using one and two Gaussian functions, respectively.

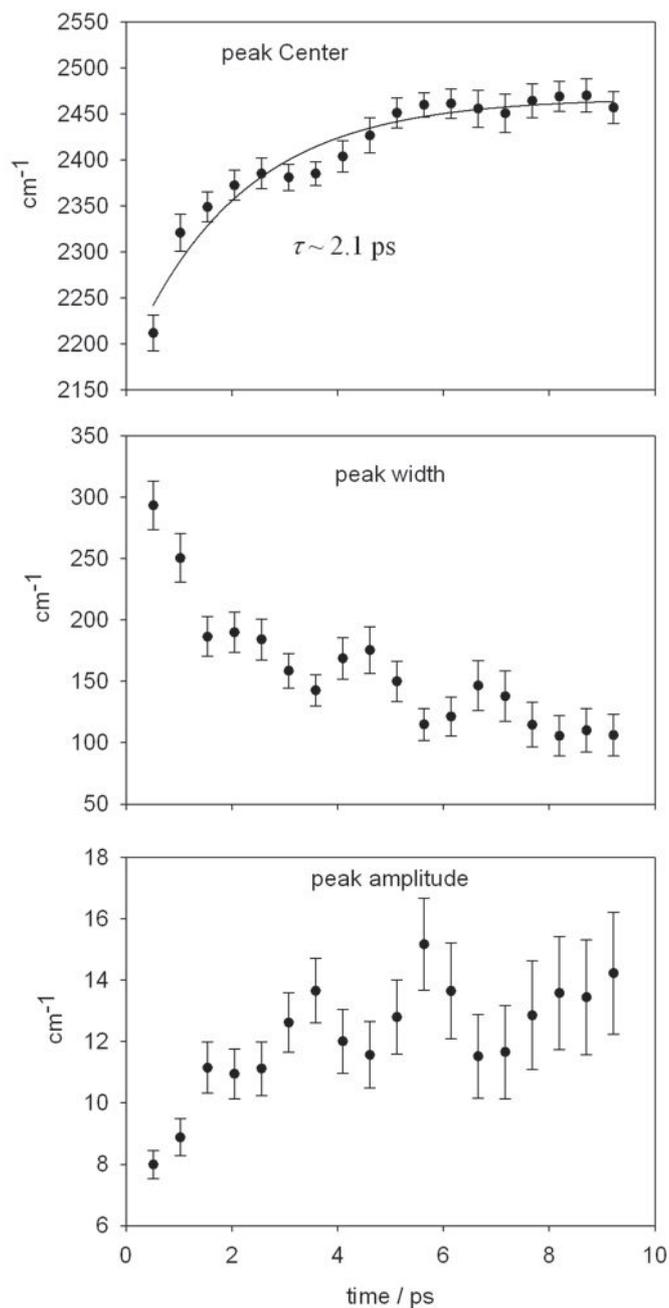

Figure 16: Results obtained from fitting the time dependent DF (relaxation) spectra with a single Gaussian. The peak centre was fit to a single exponential of the form $y = A\exp(-k_1 t) + C$

## *Transient Microsolvation Dynamics*

The DF relaxation dynamics discussed in the previous section, carried out with excited DF in an equilibrium complex with $CD_3CN$ solvent molecules, were critical in allowing us to assess the extent to which the solvent/solute potential energy function outlined in Eq (1) captures important dynamical observables across both equilibrium and non-equilibrium regimes. In order to understand the reactive dynamics, which are the ultimate aim of this work, it is important to recognize that, compared to the equilibrium solvation environment, the post-reaction solvation environment of DF is also a transient dynamical feature which contributes to its observed vibrational spectrum.

To examine the solvation environment which the DF experiences following abstraction, we carried out reactive dynamics simulations in which we damped the vibrational excitation of the nascent DF immediately following the abstraction event (i.e., at the moment of first passage through the DF equilibrium geometry), modifying the Deuterium velocity in the vibrational frame so as to remove all non-thermal vibrational excitation. This provided a set of trajectories where the initial coordinates of the solute and solvent are sampled from the distribution that follows in the immediate wake of the abstraction reaction, but with a thermal distribution of velocities for the $CD_3CN$ solvent and DF solute. The utility of these 'damped trajectories' is that they allow us to eliminate from the calculated spectra any shifts which arise as a consequence of DF vibrational excitation – thereby providing a reference baseline with which to compare spectral shifts that arise from vibrational excitation (discussed in what follows). Any time-dependence observed for DF in the damped trajectories may be assigned to relaxation of DF within the non-equilibrium solvent environment in which it finds itself immediately following the abstraction event. Fig 17 shows spectra of the DF in the picoseconds following abstraction for these damped trajectories. Fig 18 shows the results obtained by fitting the time dependent spectra to a single Gaussian function. The SI contains additional snapshots of the time dependent spectra, along with plots showing the parameters returned from the two-Gaussian fits at every snapshot. At very short times, the band is centered at 2758 $cm^{-1}$. It quickly relaxes to the equilibrium value of 2540 $cm^{-1}$ on a timescale of ~0.58 ps, a value obtained by fitting the top panel in Fig 18 to a single exponential decay. The time dependent peak width shows a profile which is similar to the peak centre: it is very broad at short times, and narrows rapidly following reaction to give a characteristic peak width around 100 $cm^{-1}$. This suggests that the DF experiences a broad distribution of solvation environments immediately following abstraction, which rapidly relax to the equilibrium limit.

Perhaps the most important conclusion to be drawn from these damped trajectories is the transience of the DF spectrum, which results from a time-dependent microsolvation environment. At very short times following abstraction, the DF has a different microsolvation environment than it does at long times; the distribution of geometries required to facilitate a reaction means that – immediately following the abstraction event – the DF has not yet had time to form intermolecular complexes with the solvent. In this sense, it feels a microsolvation environment which is somewhere between the equilibrium solvent H-bonded limit, and the gas-phase limit. This analysis allows us to deconvolute spectral shifts which arise from vibrational excitation (discussed above) and those which reflect the an effectively dynamic baseline which is linked to time-dependence in DF's microsolvation environment.

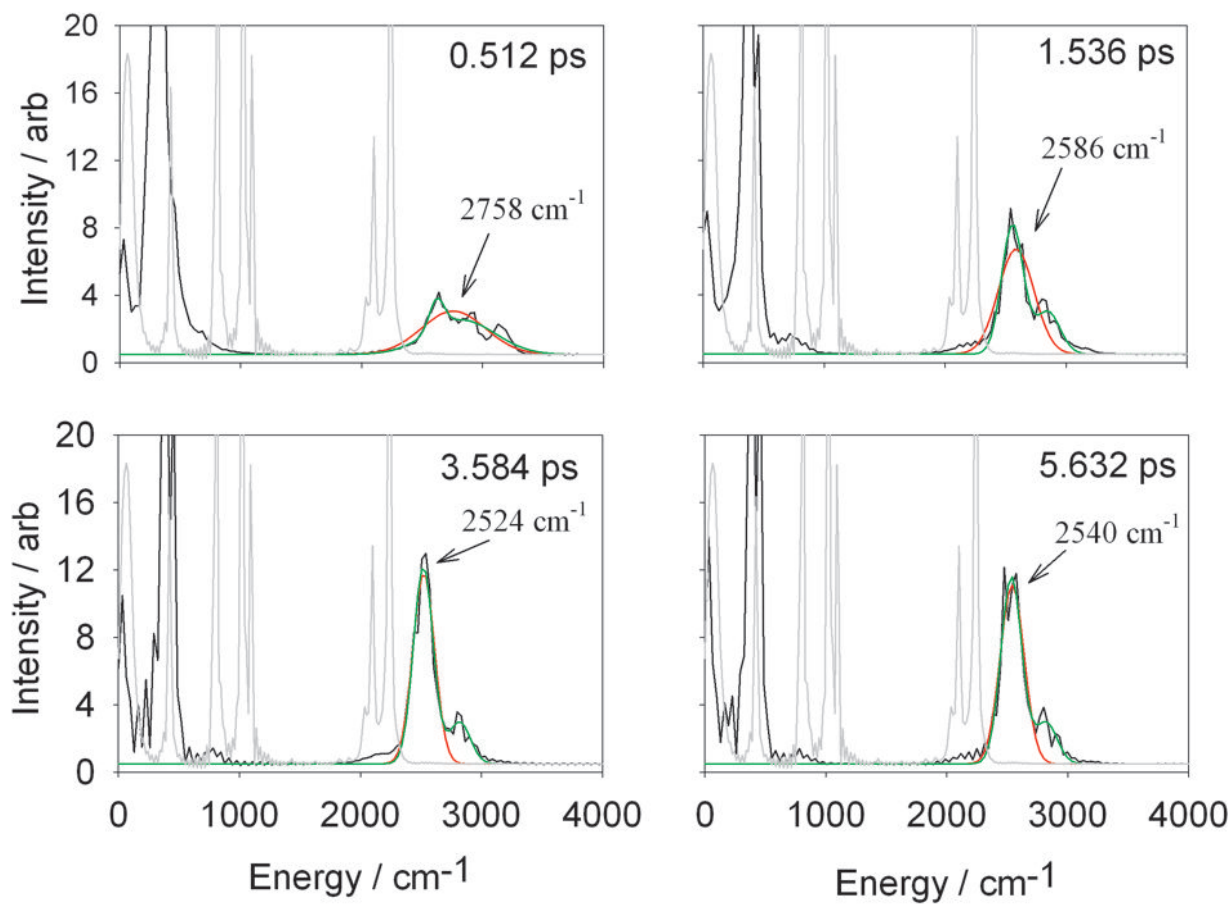

Figure 17: spectrum of DF following abstraction, obtained from the damped trajectories described above. Also indicated are the DF band centres obtained from fitting to a single Gaussian function.

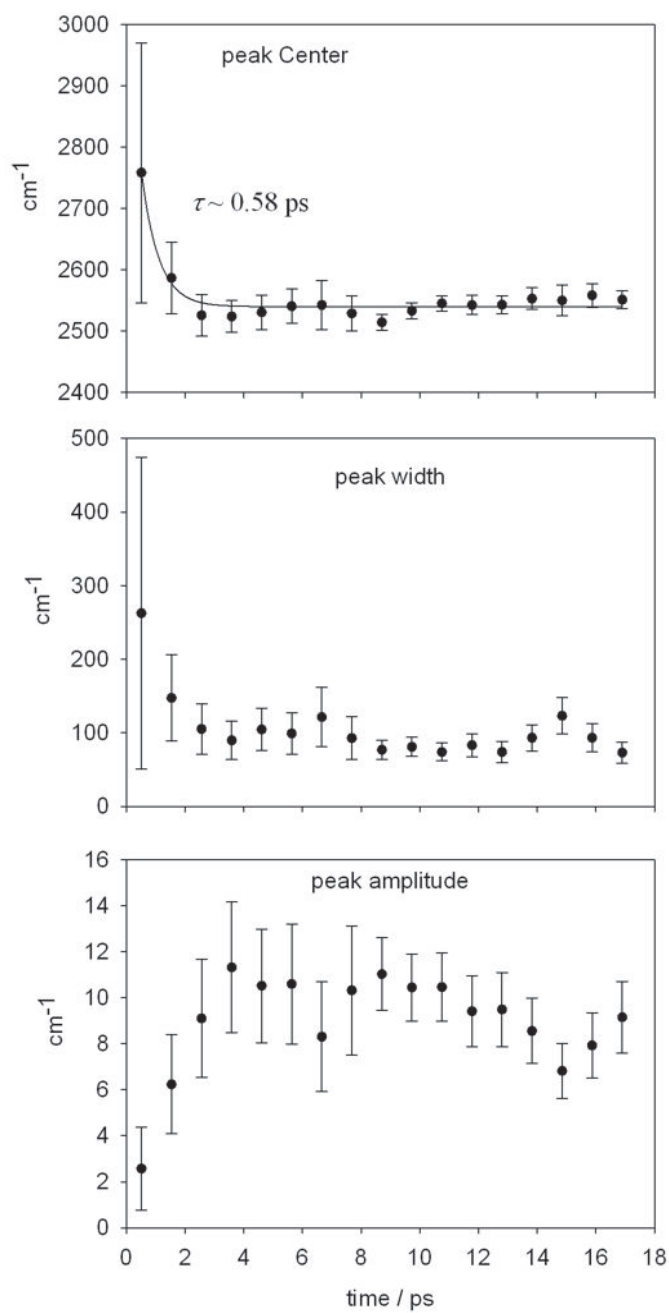

Figure 18: transient DF spectral features obtained by fitting the Fig 17 results to a single Gaussian. The peak centre has been fit to a single exponential decay

*Reaction Dynamics: Vibrational Relaxation & Transient Microsolvation*

In what follows, we discuss the results obtained from full reactive dynamics simulations, rationalizing them in terms of the results that have been presented so far. In these reactive simulations, vibrational excitation of the nascent DF arises from energy deposition following abstraction of a D atom by Fluorine from a $CD_3CN$ solvent molecule. Fig 19 shows the time-dependent DF relaxation profile obtained following the abstraction event. According to our simulations, the initial abstraction reaction deposits ~23 kcal mol$^{-1}$ vibrational energy into the DF stretch. In the harmonic approximation, one quanta of DF stretch energy corresponds to ~8.6 kcal mol$^{-1}$. Mapping this result onto our classical simulations suggests that, on average, the vibrational quantum number of the nascent DF is somewhere between 2 and 3 (i.e., 23 ÷ 8.6 ~ 2.7). Fitting the Fig 19 data with a biexponential function yields a slightly better fit with two different timescales – fast relaxation at short times with a time constant of ~1.04 ps, and a long-time relaxation rate which is slower by a factor of ~10, with a relaxation time constant of ~11.3 ps.

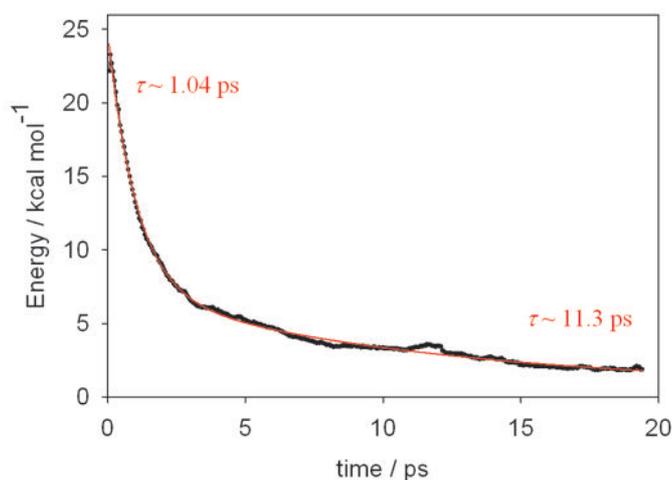

Figure 19: time-profile for vibrational energy content of DF solute in CD$_3$CN solvent following abstraction, utilizing the previously discussed 64 × 64 EVB matrix. Relaxation timescales obtained with a biexponential fit are shown in blue, while those obtained with a single exponential fit are shown in red.

Fig 20 shows the time dependent spectrum of DF following abstraction. Black lines show the DF vibrational spectra and grey lines show the equilibrium solvent spectrum. Red lines show fits to the DF stretching feature using a single Gaussian function, while green lines show fits carried out using the sum of two Gaussian functions. The SI contains additional snapshots of the time dependent spectra, along with plots showing the parameters returned from the two-Gaussian fits at every snapshot. At short times – i.e., less than a picosecond – the transient DF spectrum has a width of ~280 cm$^{-1}$, and a peak centered at 2473 cm$^{-1}$ – giving it a location which is to the red of its equilibrium position at 2540 cm$^{-1}$, and to the blue of the solvent spectrum peaks at 2101 and 2247 cm$^{-1}$. As time goes on and the DF cools, Figs 20 and 21 show that the DF band sharpens and simultaneously blue shifts toward its equilibrium position at 2450 cm$^{-1}$. This results in a decrease of both solvent/solute spectral overlap and the probability of large amplitude DF motions that strongly couple to solvent molecules. These two effects combine to decrease the energy relaxation rate from the DF solute to the CD$_3$CN solvent as time increases. The order of magnitude difference in the relaxation rates at short and long times (shown in Fig 19) is very similar to that observed in Fig 13, for DF relaxation following vibrational perturbation. The timescales for DF relaxation in Fig 19 are slower by nearly a factor of two compared to those shown in Fig 13, which is linked to the fact that the vibrationally excited DF produced from chemical reaction has poorer spectral overlap with the $CD_3CN$ solvent bands (Fig 20) than DF whose initial excitation arises from vibrational perturbation (Fig 15).

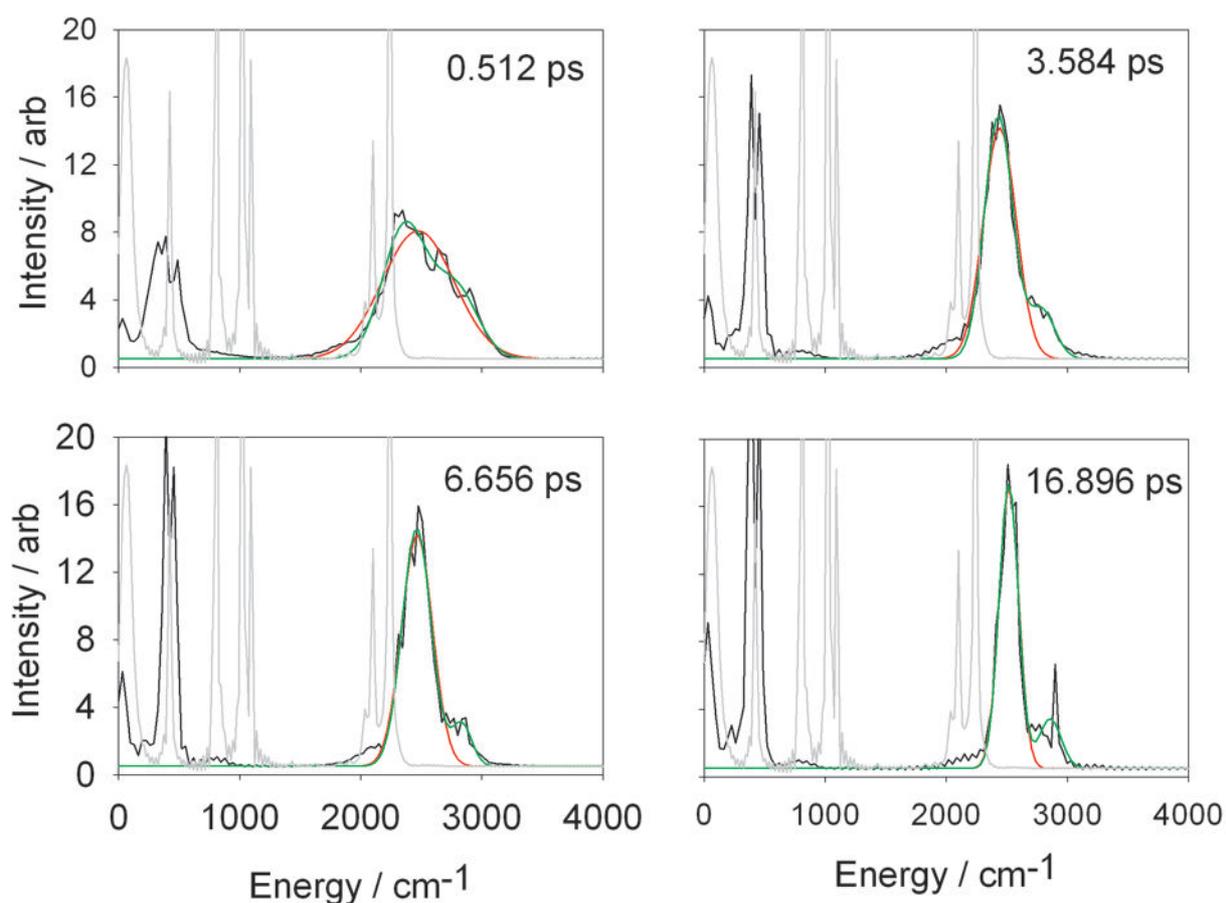

Figure 20: time dependent spectra of nascent DF following reaction with $CD_3CN$

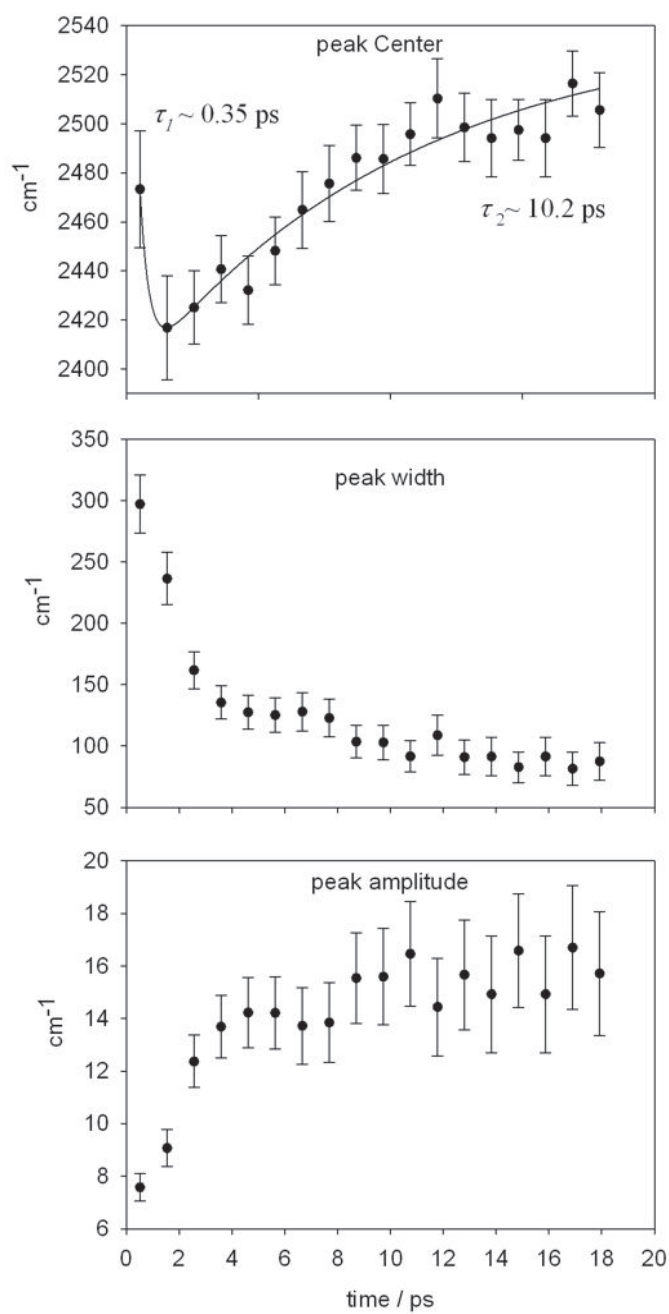

Figure 21: Results obtained from fitting the time dependent DF (reactive) spectra with an single Gaussian; the fits are obtained from a function of the form $y = A\exp(-k_1 t) + B\exp(-k_2 t) + C$, where $C$ is constrained to 2540 cm$^{-1}$.

The time profile of the post-reaction spectral data shown in Figs 20 and 21 varies substantially from the relaxation spectral data in Figs 15 and 16. For the relaxation data, the transient spectral features are easy to rationalize: as a result of vibrational excitation, there is a prompt red shift in the DF band, from 2211 cm$^{-1}$ ← 2540 cm$^{-1}$. This shift of 329 cm$^{-1}$ is close to the shift of ~300 cm$^{-1}$ which we would expect from gas phase simulations of a Morse oscillator (Table 3). Following this prompt red shift, there is a gradual blue shift back to the equilibrium band centre (2540 cm$^{-1}$) as the excited DF loses its vibrational energy.

The position of the DF spectral peak as a function of time in Fig 16 is well represented using a single exponential function. For vibrationally excited DF produced following chemical reaction, the DF spectral position in Fig 21 shows a rather more complicated time profile, with two important differences: (1) the initial prompt red shift gives a DF band centered at 2473 cm$^{-1}$, far less deep into the red than the initial value of 2211 cm$^{-1}$ in Fig 16; and (2) the DF band centre goes through a transient red shift at short times, taking it through a minimum of 2416 cm$^{-1}$ before blue shifting back toward equilibrium. To better understand these timescales, the results in Figure 21 were fit using a biexponential function of the form

$$\omega_0(t) = A\exp(-k_1 t) - B\exp(-k_2 t) + C \qquad (16)$$

Understanding this complicated time dependence requires recognition that – in the case of the reactive dynamics – the spectral baseline with respect to which shifts occur is not constant, on account of the fact that the microsolvation environment of the DF is time-dependent, as discussed in the previous section.

Fig 22 shows a time series of radial distribution functions (RDFs) between the D atom in DF, and the N atoms in the CD$_3$CN solvent molecules, and provides insight into solvent structural differences that accompany the relaxation following reaction versus that which follows vibrational perturbation in an equilibrium solvation environment. For the RDFs obtained following DF vibrational excitation in an equilibrium solvation environment, there is a distinct shift in RDF peak position with time, from 1.5 Angstroms at short times to 1.65 Angstroms at long time. This well-defined shift is consistent with DF remaining complexed to solvent molecules during the relaxation process, and is easy to rationalize as a consequence of anharmonicity in the DF oscillator: at high vibrational energies, the average DF bond length is longer, which corresponds to a smaller CD$_3$CN–DF distance. Consistent with this hypothesis is the fact that the time-dependent RDFs are essentially identical beyond distances of ~2 Å. The RDFs obtained following reaction have a rather different profile. The distinguishing feature is not a peak shift, but rather the width of the distribution: at short times, the distribution is considerably wider than at longer times, with substantial amplitude at distances larger than 2 Å. This provides strong evidence for the fact that the nascent DF created following an abstraction reaction sees a wide range of microsolvation environments beyond the solute/solvent complexes that characterize the equilibrium RDF.

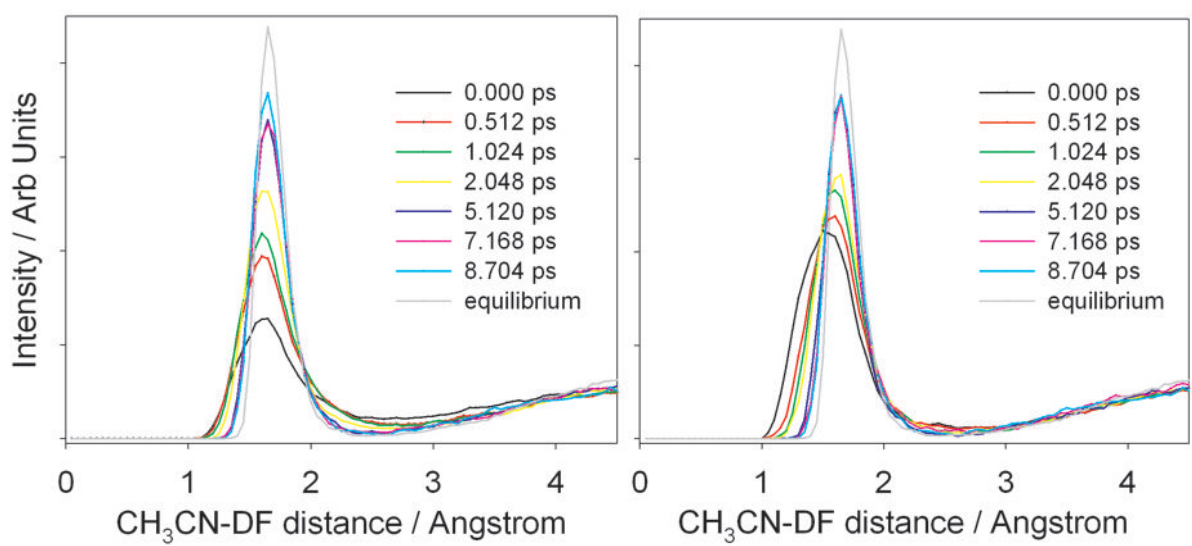

Figure 22: Time dependent RDFs obtained following DF relaxation. The left hand panel shows the RDF obtained when DF* is produced from the reactive dynamics; the right hand panel shows the RDF obtained when DF* relaxes within an equilibrium solvation environment.

Fig 23 synthesizes all of the results in the previous sections, in order to provide insight into the time profile of the DF band centre in Fig 21. Immediately following reaction, the microsolvation environment felt by the DF is intermediate between the gas phase and the equilibrium (complexed) values, with a baseline of ~2758 cm$^{-1}$. Vibrational excitation of the DF, as a result of chemical reaction, results in a prompt red shift of approximately 285 cm$^{-1}$, giving a vibrationally excited DF band centre at ~2473 cm$^{-1}$. The magnitude of this prompt red shift, illustrated in Fig 23 with a red arrow, is very similar to that observed in both DF relaxation simulations, and that which we would predict for a gas phase Morse oscillator (Table 3). The prompt shift, illustrated with a red arrow in Fig 23, takes DF less far into the red than occurs for the relaxation dynamics (2211 ← 2540 cm$^{-1}$, also shown in Fig 23), but farther into the red than occurs for the gas phase dynamics (2707 ← 2998 cm$^{-1}$, Table 3). The origin of these differences concerns the spectral baseline of the DF stretch prior to any vibrational excitation: in the gas phase, the baseline band centre is 2998 cm$^{-1}$; in the complexed relaxation dynamics, the baseline centre is 2540 cm$^{-1}$; and in the reactive case, the baseline centre is 2758 cm$^{-1}$. Following the prompt red shift, there are then two competing effects. First, there is a fast red shift that occurs as the DF solvation environment relaxes – i.e., undergoing rotational and translational diffusion to form H-bonded complexes with neighboring solvent molecules. The time constant for this shift is on the order of 0.35 ps, in good agreement with the transient microsolvation timescales seen in the damped trajectories (Fig 18). On top of this red shift, there is a blue shift (illustrated in Fig 23 with blue arrows) on account of DF losing its vibrational energy to the solvent, which has a time constant on the order of 10.2 ps, in reasonable agreement with long time DF relaxation timescale of 11.3 ps shown in Fig 19. The time-dependent spectral profile of the nascent DF formed following reaction, shown in Fig 21 and also in Fig 23, is the result of these opposing effects, attributable to the opposing spectral effects of solvent environment relaxation and vibrational relaxation.

For contrast, Fig 23 also shows the results obtained from DF relaxation following vibrational perturbation, where the DF microsolvation environment is essentially time-independent – i.e., DF remains complexed to the solvent molecules throughout. In this case, the spectral baseline is constant, and corresponds to the equilibrium DF spectrum shown in Fig 14. The prompt vibrational excitation can be imagined to shift the baseline toward the red at time zero, and subsequent vibrational relaxation of tbhe DF results in a transient blue shift back to equilibrium. In this case, where there is little time dependence in the spectral baseline, the transient DF vibrational spectral profile is much easier to understand.

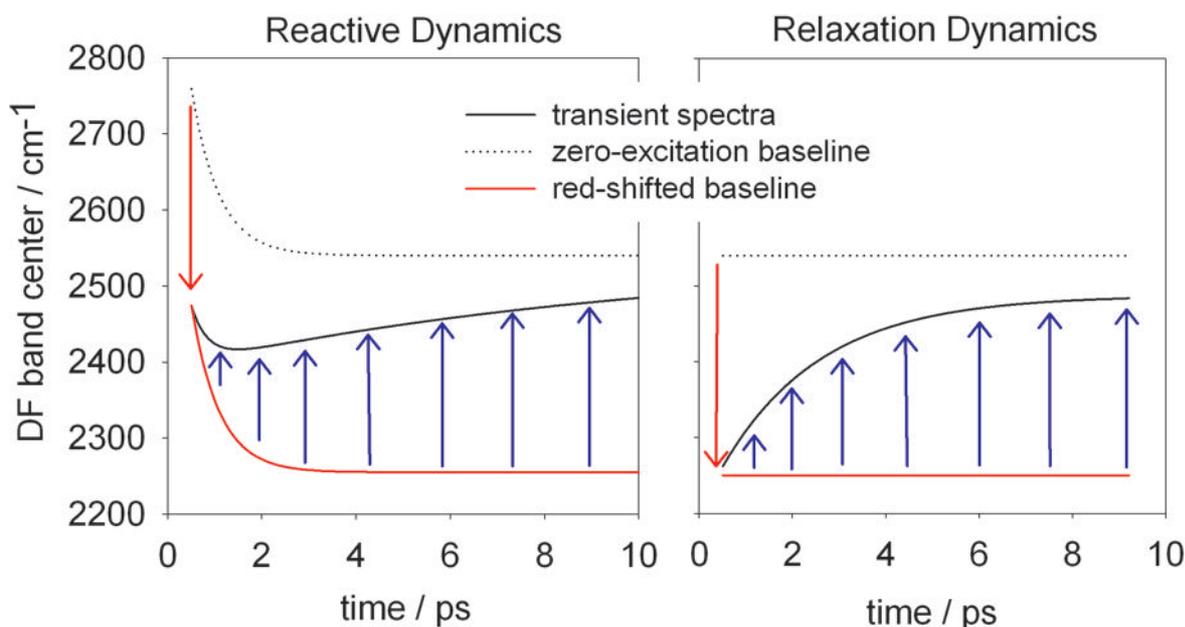

Figure 23: schematic diagram illustrating the contrasting DF spectral time profiles (black line) seen in the reactive dynamics versus the relaxation dynamics (i.e., the data in Figs 21 and 16, respectively). The dotted line illustrates the spectral baseline (i.e., position of the center) without any DF excitation, and the red solid line corresponds to the spectral baseline red shifted according to the prompt DF vibrational excitation at time zero (red arrow). The blue arrows illustrate the blue shift that occurs as DF relaxes to equilibrium, losing its vibrational energy to the solvent.

## Conclusions

In this work, we have described in detail an MPI-parallelized multi-state valence bond implementation within the CHARMM program suite (and also within TINKER[27]) which allows one to construct reactive force fields using multi-state molecular mechanics approaches for carrying out non-equilibrium MD simulations. Conveniently, it allows one to use any of the force field approaches and various utilities which are available within general MD packages like CHARMM and TINKER, along with a range of different functional forms for describing the coupling elements.

As an initial application of this multi-state framework, and in order to demonstrate the sorts of simulations and analyses it can be used to carry out, we have investigated non-equilibrium reaction dynamics of F + $CD_3CN$ abstraction reactions in $CD_3CN$ solvent.[66] The approach we outline herein has a number of satisfactory features. It is able to reproduce the $CD_3CN$ experimental solvatochromatic spectral shift of DF compared to its gas phase spectrum, as well as shifts related to DF vibrational excitation. The energy relaxation timescales for excited DF are also in excellent agreement with those observed experimentally.[66] Analysis of the detailed transient DF spectra provide additional microscopic insight, and suggest that the phenomenological spectra observed during reaction dynamics experiments result from two competing effects: a blue shift linked to vibrational relaxation, and a red shift linked to relaxation of the DF microsolvation environment.

This work, aimed at investigating reaction dynamics in strongly coupled solvents which H-bond to the nascent solute, establishes an important limit that complements our previous studies of solution-phase reactions dynamics in weakly coupled solvents. The results show that – despite strong coupling – vibrational excitation of the nascent products persists for an appreciable timescale. It will be fascinating to explore the extent to which persistent vibrational excitation of the sort observed herein impacts reaction outcomes in more complex chemical systems. For example, there is evidence that product branching ratios for reactions in thermal synthetic chemistry as well-known as alkene hydroboration are sensitive to the both the extent of initial vibrational excitation, and the solvent/solute coupling that governs its subsequent dissipation.[27, 71]

The framework that we have used to carry out the simulations described in this article is entirely general, insofar as it can be used to describe arbitrary reactive systems. It scales linearly with the number of available CPU cores, and is designed to exploit massively parallel computational architectures. The increasing interest in the study of condensed phase dynamical systems along with the poor scaling of electronic structure theory approaches suggests that the use of methods like MS-EVB will remain widespread, given their balance between accuracy and efficiency. In future work, it will be fascinating to explore how such methods scale to massive parallel architectures, and whether it is possible to build MS-EVB models by fitting to on-the-fly electronic structure theory using more sophisticated strategies. It will also be interesting to investigate the extent to which the accuracy of the reactive potentials can be improved by building the diabatic states from polarizable force fields, and whether such treatments provide any further insight into reaction and relaxation processes in solution.

# Supplementary Information

This Supplementary Information includes:

(1) Time dependent spectra of DF following a non-equilibrium vibrational perturbation;

(2) Results obtained from fitting the DF relaxation spectra to two Gaussians;

(3) Time dependent spectra of DF obtained from the damped trajectories described in the main text

(4) Results obtained from fitting the damped trajectory spectra to two Gaussians;

(5) Time dependent spectra of DF obtained from full reactive dynamics simulations;

(6) Results obtained from fitting the reactive dynamics spectra to two Gaussians.

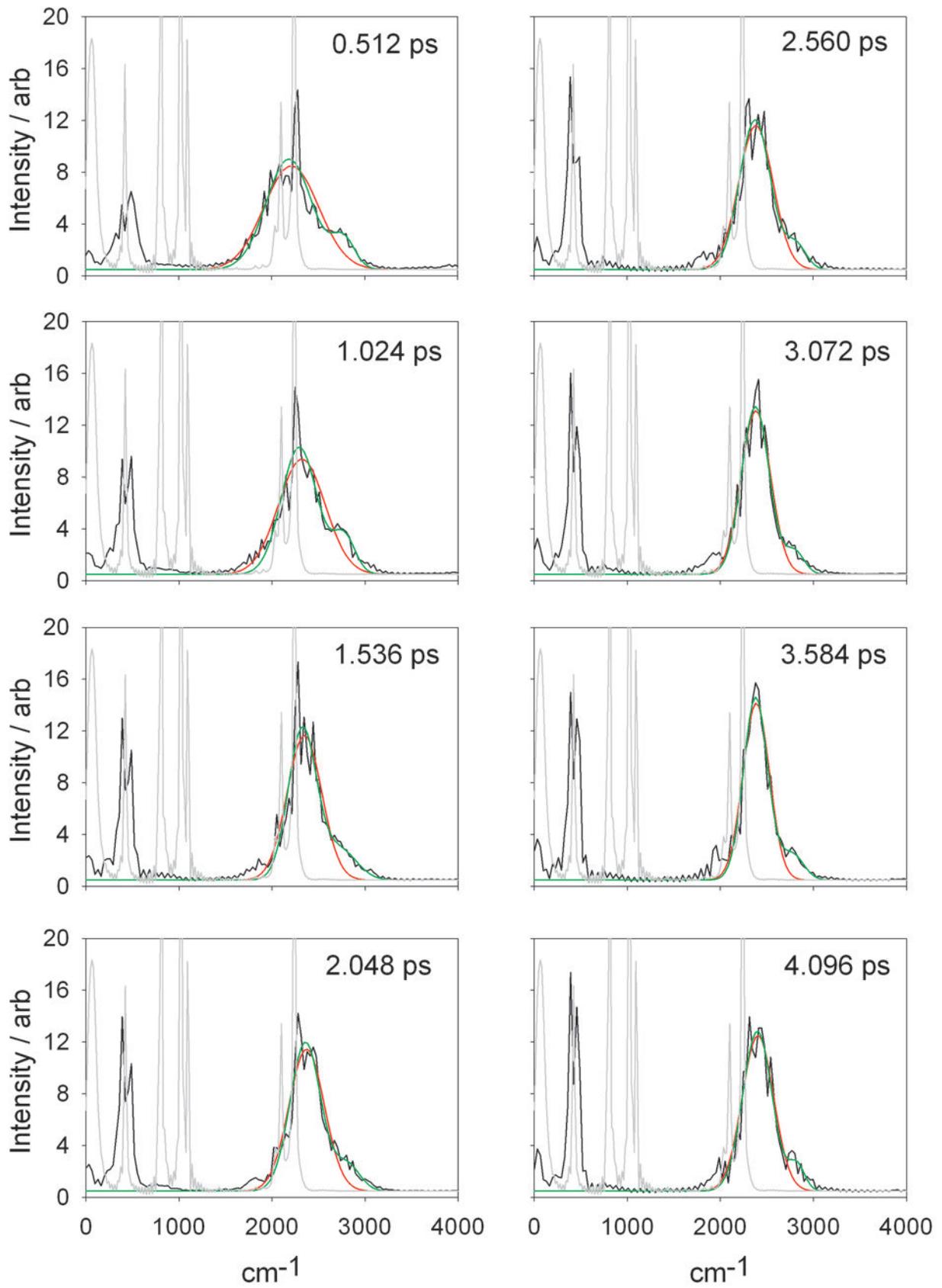

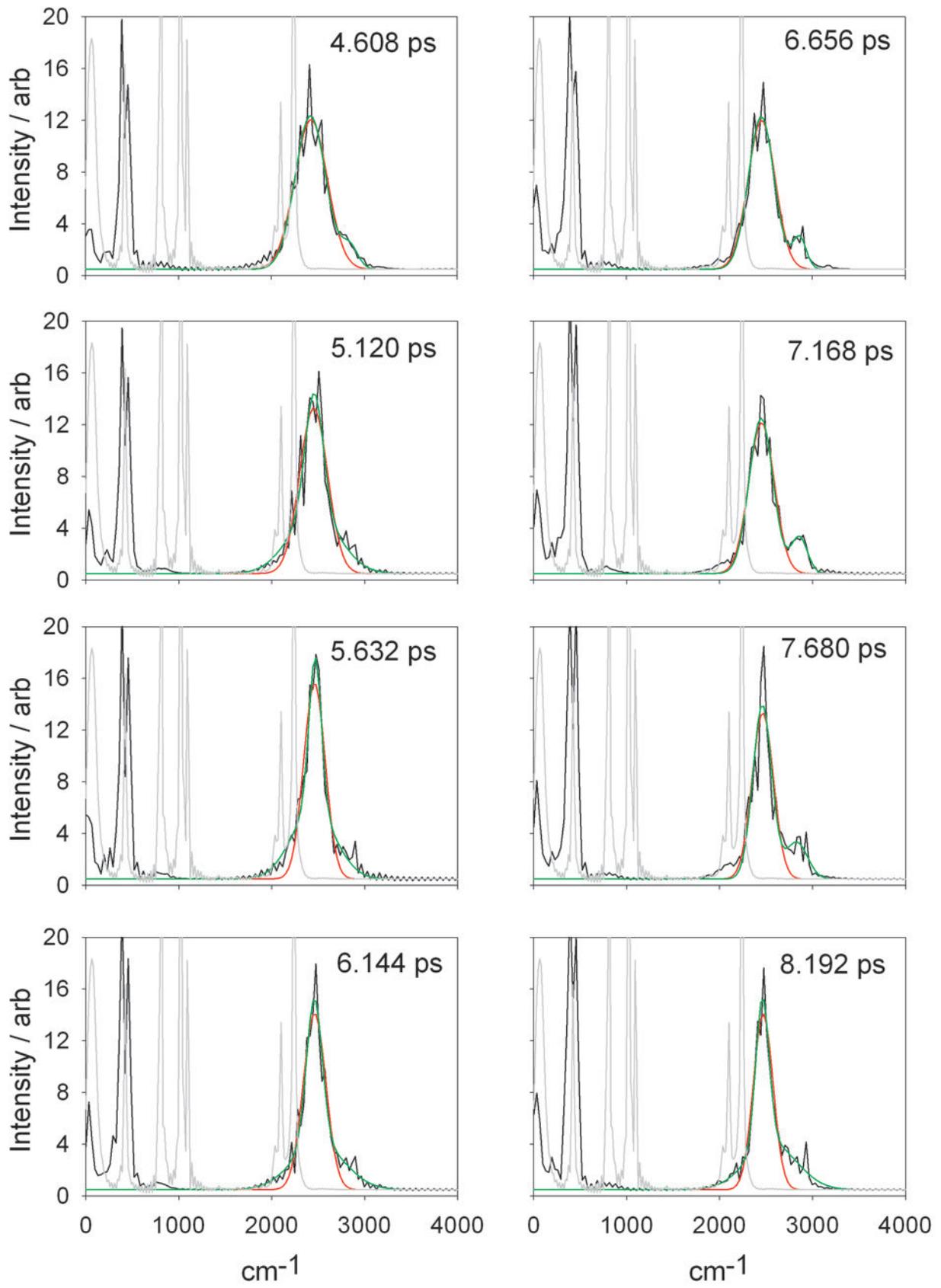

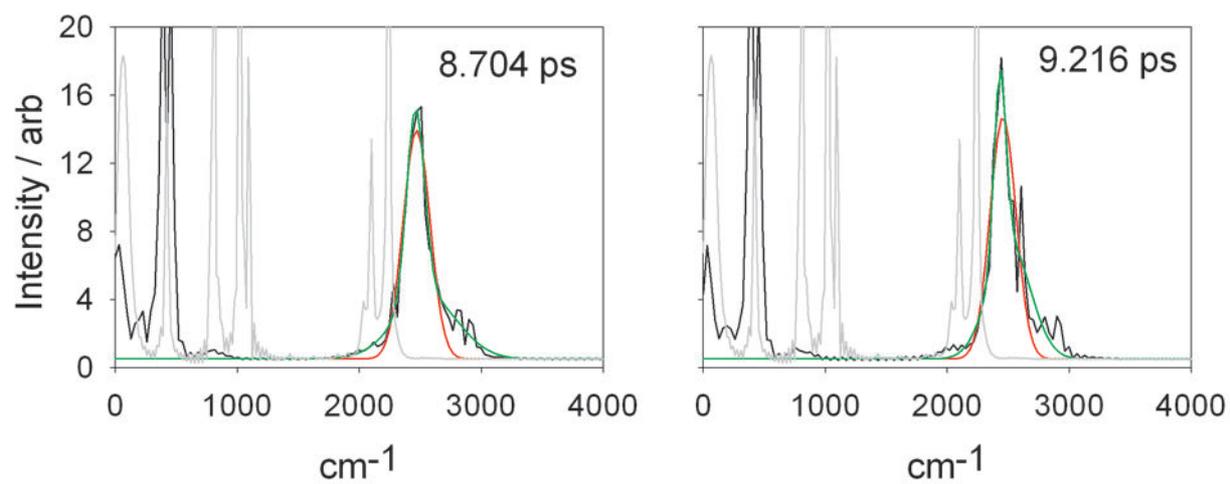

Figure S1: time dependent relaxation spectra of DF following a non-equilibrium 'kick', showing the $CD_3CN$ vibrational spectrum (grey) and the DF vibrational spectrum (black). The red and green lines are fits to the DF vibrational spectrum, using one and two Gaussian functions, respectively.

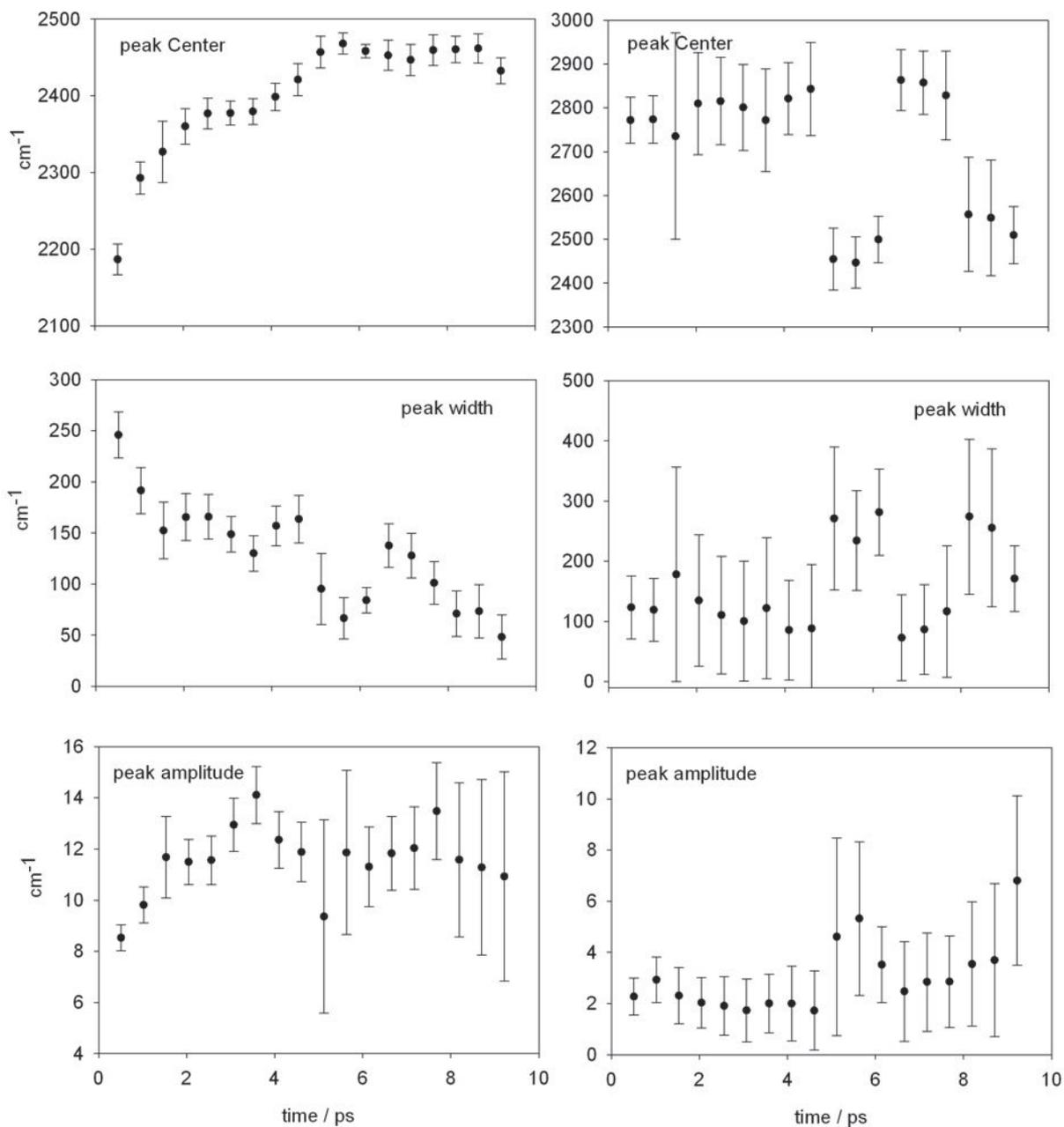

Figure S2: Results obtained from fitting the time dependent DF (relaxation) spectra in Fig S1 with two Gaussians. Left and right panels correspond to fit parameters obtained for the first and second Gaussian function, respectively.

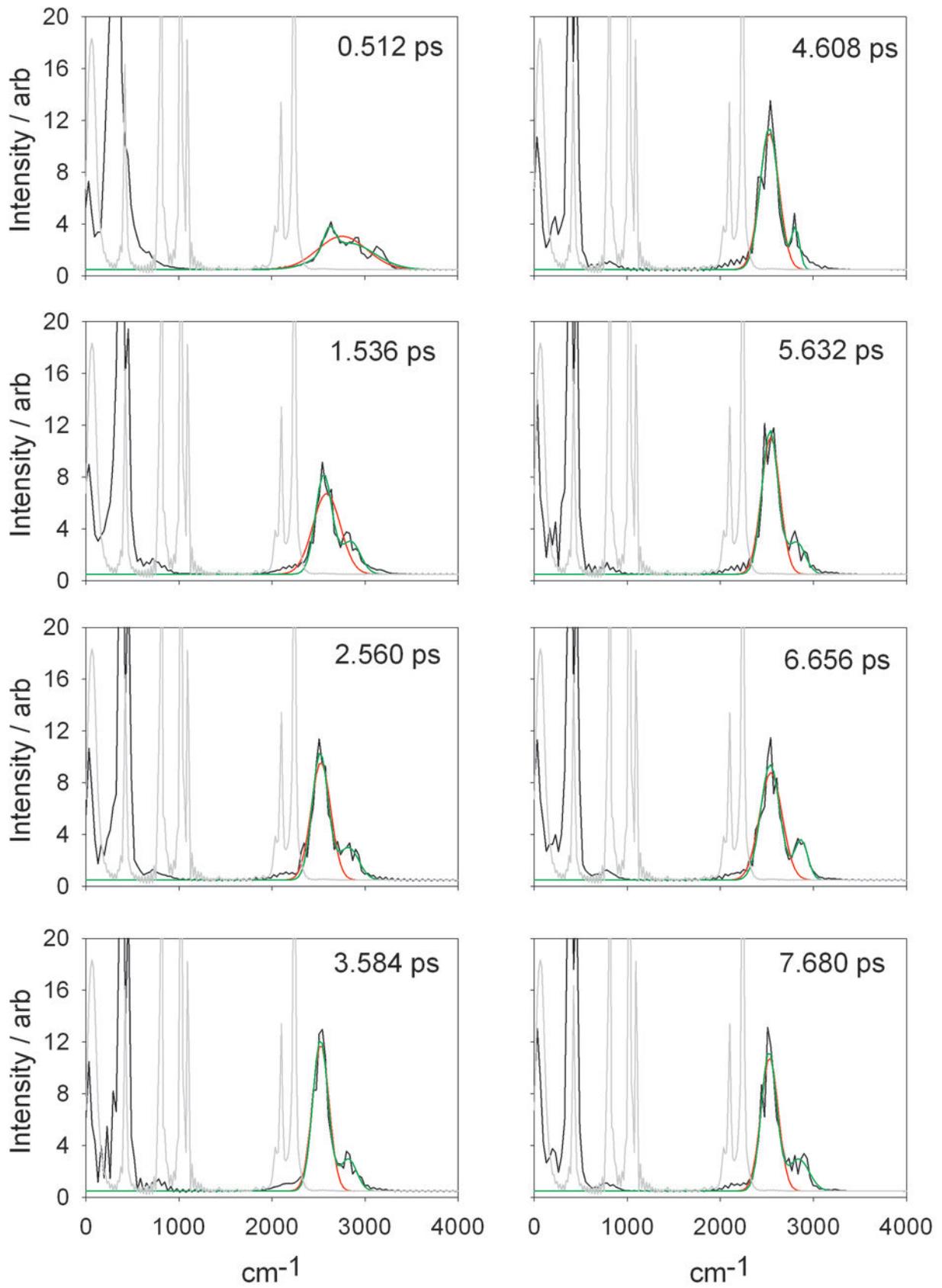

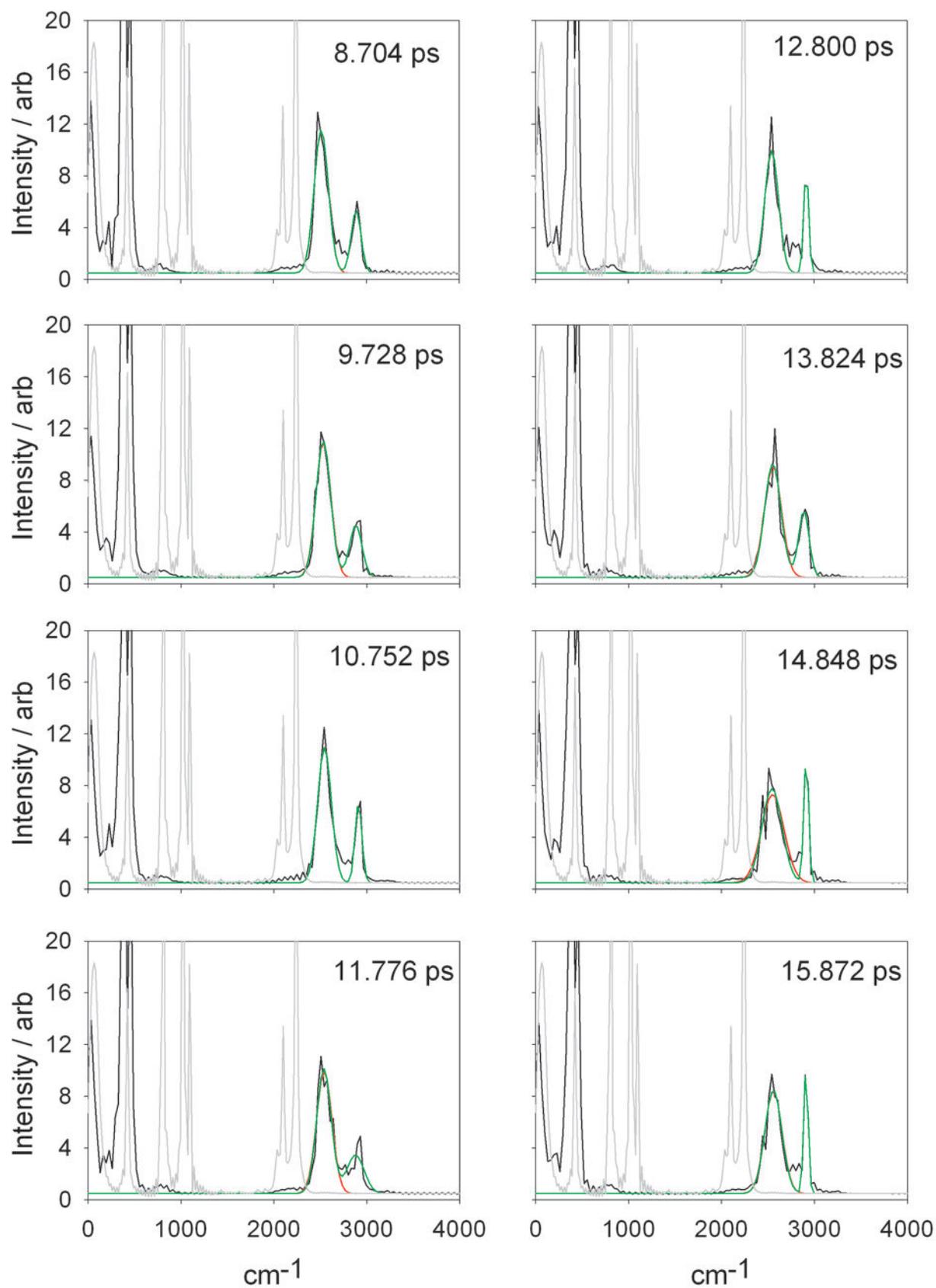

Figure S3: spectrum of DF following abstraction, obtained from the damped trajectories', showing the $CD_3CN$ vibrational spectrum (grey) and the DF vibrational spectrum (black). The red and green lines are fits to the DF vibrational spectrum, using one and two Gaussian functions, respectively.

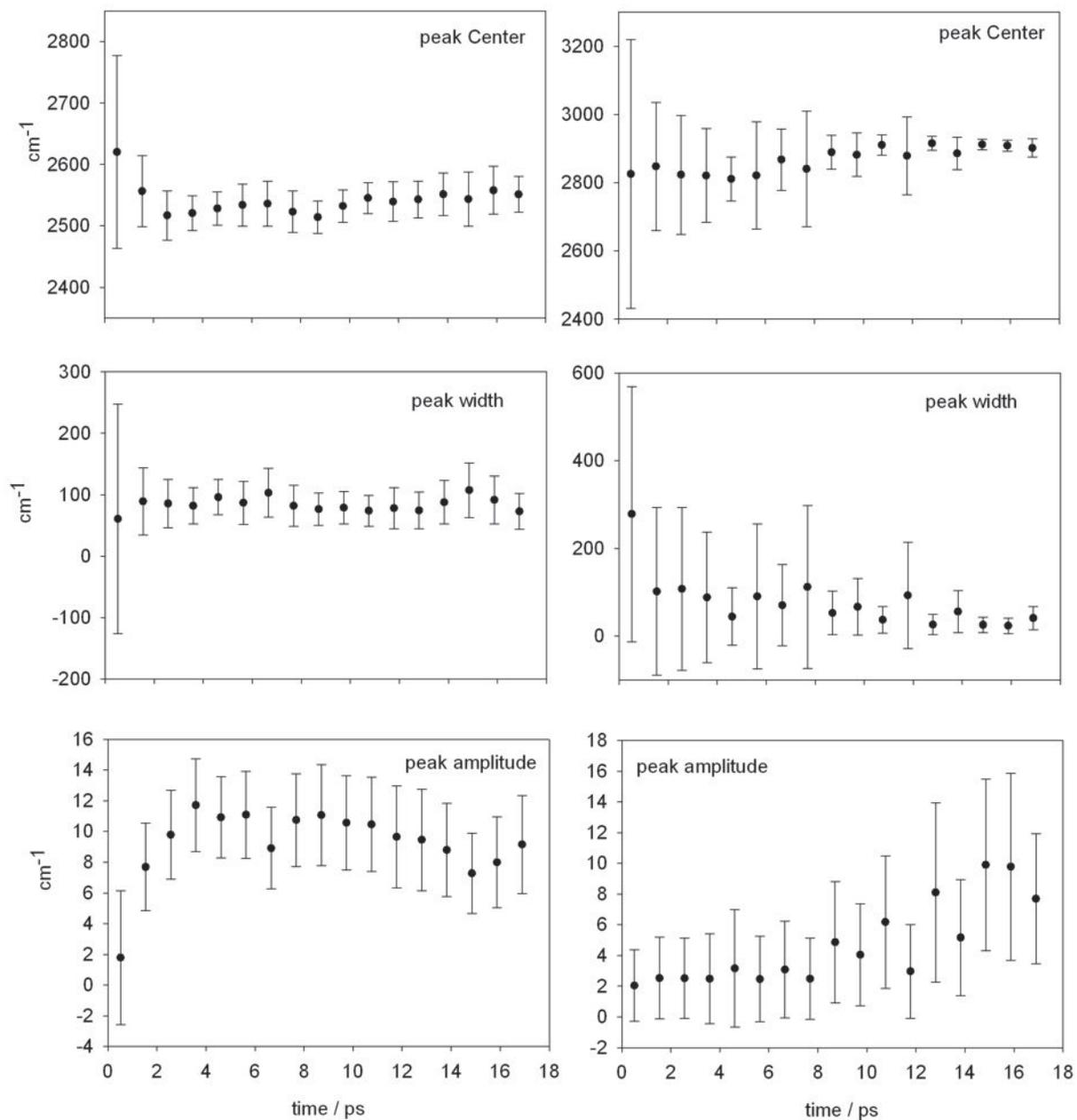

Figure S4: Results obtained from fitting the damped trajectory DF spectra in Fig S3 with two Gaussians. Left and right panels correspond to fit parameters obtained for the first and second Gaussian function, respectively.

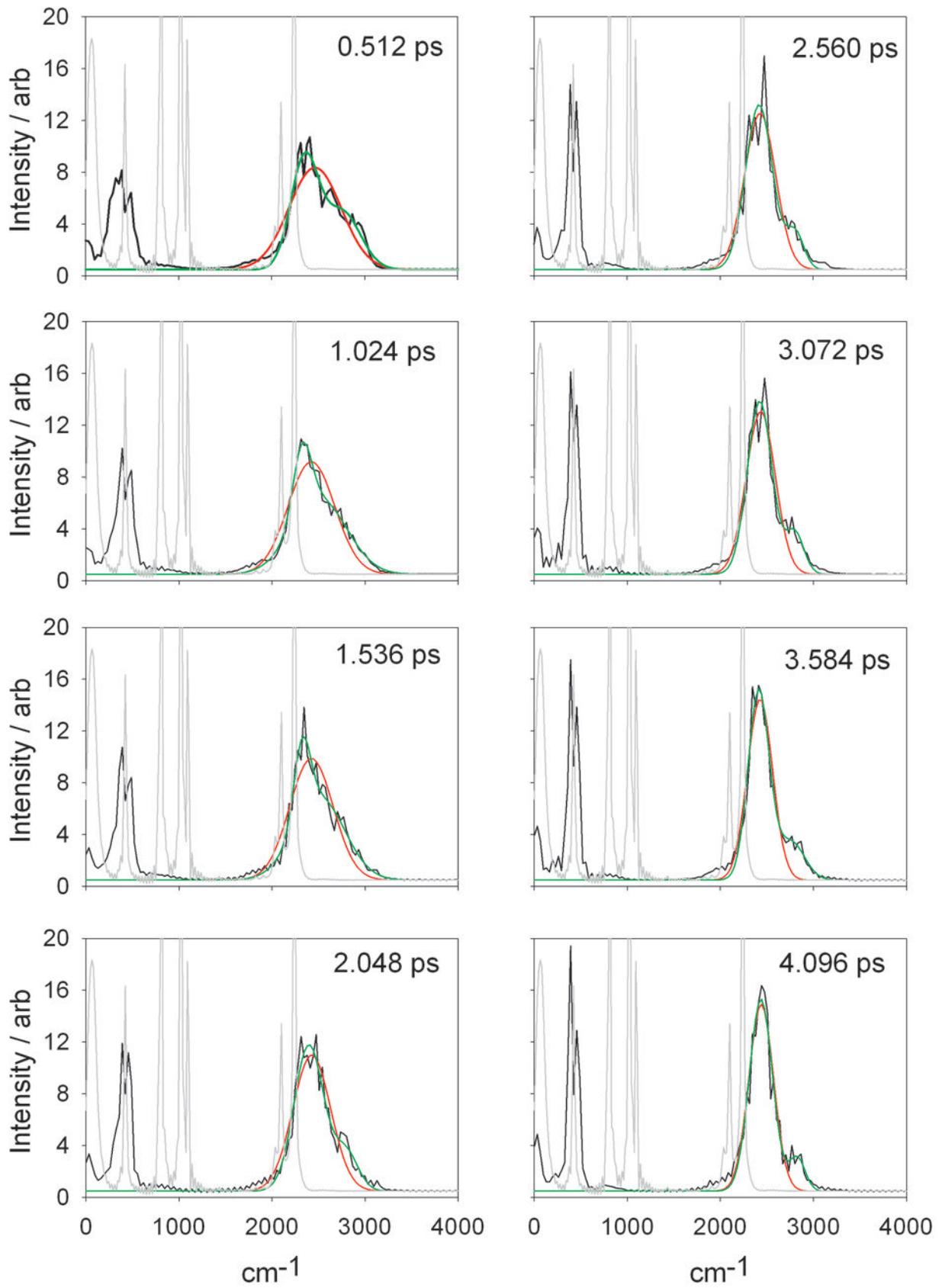

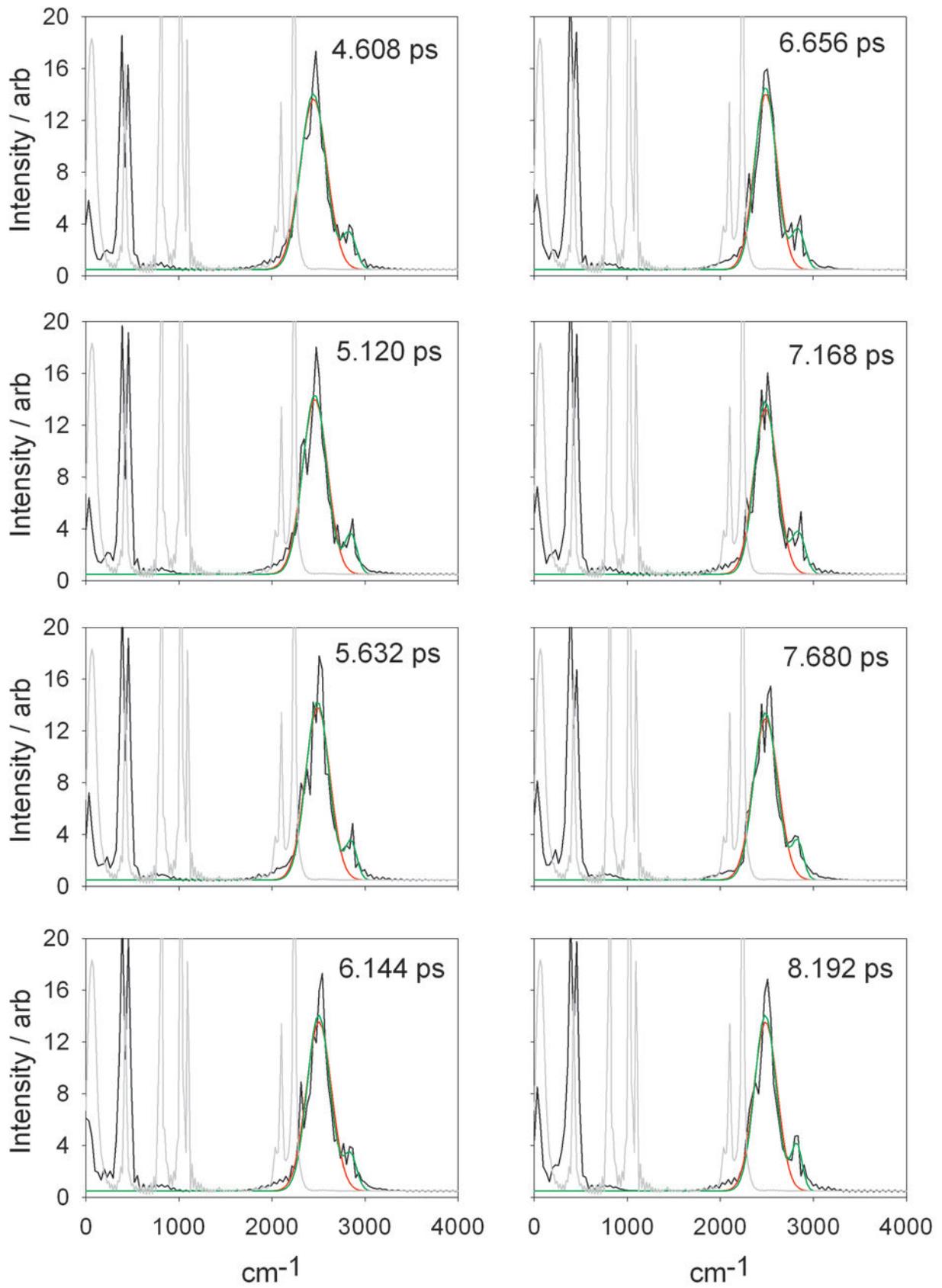

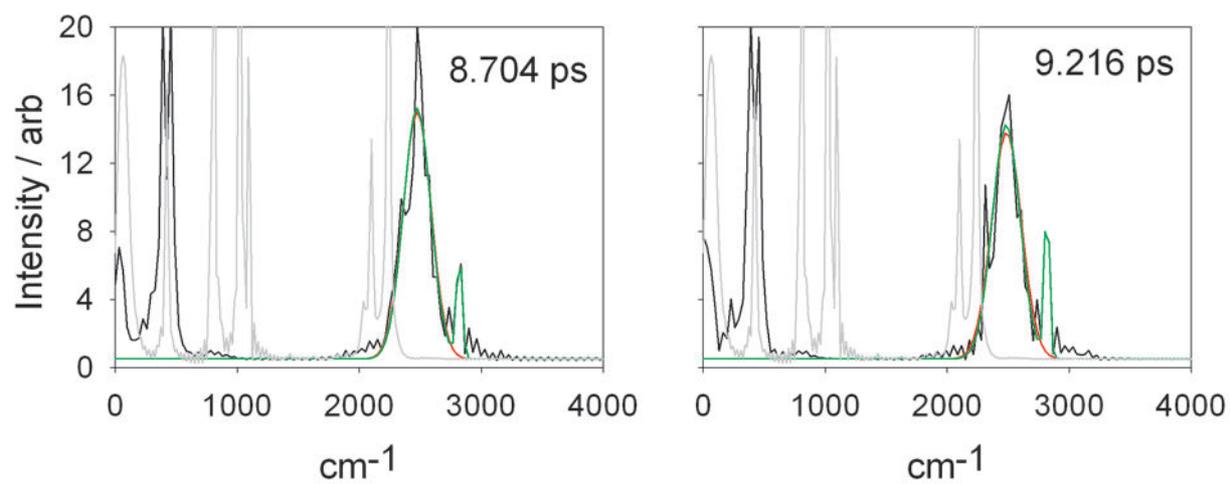
Figure S5: time dependent spectra of nascent DF following reaction with CH₃CN

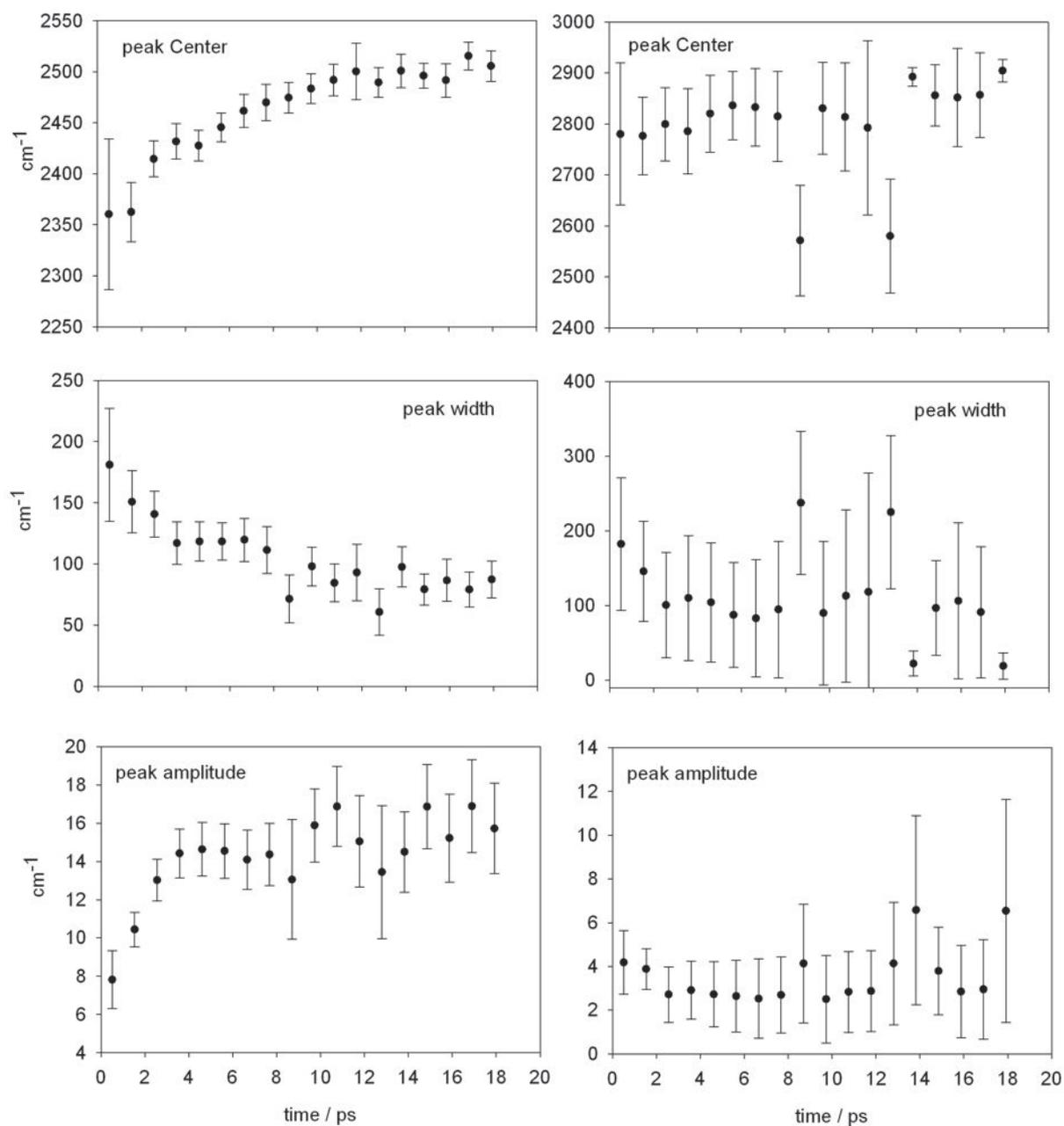

Figure S6: Results obtained from fitting the time dependent DF (reactive) spectra with two Gaussians. Left and right panels correspond to fit parameters obtained for the first and second Gaussian function, respectively

## Acknowledgements

DRG acknowledges support from the Royal Society as a University Research Fellow. AJOE recognizes financial support from the Engineering and Physical Sciences Research Council (EPSRC, Programme Grant EP/G00224X and a studentship for GTD) and the European Research Council (ERC, Advanced Grant 290966 CAPRI). JNH acknowledges a Royal Society Wolfson Merit Award.